# Comprehensive Survey of Ternary Full Adders: Statistics, Corrections, and Assessments

Sarina Nemati [1], Mostafa Haghi Kashani [2], Reza Faghih Mirzaee [2*]

[1] Faculty of Computer Science and Engineering, Shahid Beheshti University, G.C., Tehran, Iran
[2] Department of Computer Engineering, Shahr-e-Qods Branch, Islamic Azad University, Tehran, Iran

[*] Corresponding Author's E-Mail: r.f.mirzaee@qodsiau.ac.ir

**Abstract:** The history of ternary adders goes back to more than six decades ago. Since then, a multitude of ternary full adders (TFAs) have been presented in the literature. This paper conducts a review of TFAs so that one can be familiar with the utilized design methodologies and their prevalence. Moreover, despite numerous TFAs, almost none of them are in their simplest form. A large number of transistors could have been eliminated by considering a partial TFA instead of a complete one. According to our investigation, only 28.6% of the previous designs are partial TFAs. Also, they could have been simplified even further by assuming a partial TFA with an output carry voltage of 0V or $V_{DD}$. This way, in a single-$V_{DD}$ design, voltage division inside the Carry generator part would have been eliminated and less power dissipated. As far as we have searched, there are only three partial TFAs with this favorable condition in the literature. Additionally, most of the simulation setups in the previous articles are not realistic enough. Therefore, the simulation results reported in these papers are neither comparable nor entirely valid. Therefore, we got motivated to conduct a survey, elaborate on this issue, and enhance some of the previous designs. Among 84 papers, 10 different TFAs (from 11 papers) are selected, simplified, and simulated in this paper. Simulation results by HSPICE and 32nm CNFET technology reveal that the simplified partial TFAs outperform their original versions in terms of delay, power, and transistor count.

## I. Introduction

Multiple-valued logic (MVL) is a computational approach where there are more than just two truth values. In a more general sense, Fuzzy logic, one of whose intentions is to represent uncertain information, is an extension of MVL to infinite possibilities. Both concepts are in contrast with the traditional binary logic, which is founded upon dualism. Therefore, MVL systems are more compatible with real-world situations and human needs. Jan Lukasiewicz created the first MVL system in 1920. Since then, MVL and Fuzzy logic have been viable alternatives to traditional systems. Nowadays, Fuzzy logic is widely used in natural language processing (NLP) and many other artificial intelligence applications [1, 2]. Fuzzy sets are also broadly used to describe imprecise or vague data [1].

Additionally, at the hardware level, MVL on-chip buses have been proved to be promising [3, 4]. What they do is facilitate the transmission of more information over a bus line. Furthermore, triple- and quad-level cell (TLC and QLC) flash memories are currently being used in industry [5]. They mostly benefit from high data density and low costs. Finally, quantum devices and computers are inherently multivalued [6].

Addition is one of the four basic arithmetic operations. Lying on the critical path, an adder is responsible for almost every calculation inside an arithmetic logic unit (ALU). As a result, its improvement enhances the performance of the entire system. The design of ternary half and full adders (THAs and TFAs) has been a popular subject over the past years. Circuit designers have deployed various logic styles and technologies to reach high-speed, low-power, area-efficient THAs and TFAs. Speed, power consumption, and area are the three main evaluating factors in very large-scale integration (VLSI) design.

Transistors, including bipolar junction transistors (BJTs) and field-effect transistors (FETs), have always been inseparable electronic components from analog and digital circuits. Since there are more than two voltage levels in MVL, multi-threshold transistors are essentially needed for the detection of different voltage levels. Although multi-threshold CMOS (MTCMOS) exists [7], it does not provide the high flexibility required for MVL circuit design. After the advent of carbon nanotube FETs (CNFETs) [8], MVL circuitry has soared in popularity because the threshold voltage of the CNFET device is adjustable by the diameter of carbon nanotubes (CNTs). Most of the literature's ternary and quaternary arithmetic circuits are based on this emerging nanoscale technology.

There are only few brief reviews of the design of ternary logic circuits in the literature [9, 10]. However, to the best of our knowledge, there is not any particular review of ternary adders. We got motivated to undertake a review of TFAs mainly because:

1. There is not any review of TFAs, although there are plenty of published papers about them.
2. The given TFAs in the previous articles have been compared to a limited number of competitors. Hence, the efficiency level of the presented TFAs compared to a wider range of competitors is unknown.
3. The simulation setup in one paper is different from the simulation environment in others. For example, TFAs have been simulated by using different input waveforms. It is impossible/unfair to compare the results when the conditions are not equal.
4. Real simulation test-beds have been neglected in most of the previous papers. For instance, almost all prior TFAs have been simulated without employing any input buffer or a reasonable output load. Therefore, the simulation results reported in these papers are not entirely valid. Besides, a complete input pattern is required to estimate maximum cell delay precisely; however, an incomplete input pattern, missing a large number of transitions, has been used in the previous works. Thus, it is uncertain whether the reported values are truly representative of cell delay or not.

Furthermore, ternary adders usually have a large number of transistors. Circuit designers have always targeted a design with fewer transistors. However, most of the previously presented TFAs are not in their simplest form. The authors could have eliminated more transistors and simplified their designs to a further degree. This has been the second motivation for writing this paper to make a few corrections to some of the previous designs. The following is the synopsis of our main objectives for writing this paper:

1. To obtain statistical information about the number of TFAs whose structures are not in their simplest form. This number can reveal how challenging the design of a TFA might be despite its apparent simplicity.
2. To show how some of the TFAs whose structures are not in their optimal form can be simplified.
3. To simulate TFAs in a realistic simulation test-bed and achieve more accurate results.

Also, we are going to answer the following main research questions:

1. Which design methodologies have previously been used in the literature, and with what prevalence?
2. Are the prior designs in their simplest form? If not, how can we simplify them even further?
3. What are the unresolved issues?

The rest of the paper is organized as follows: A statistical and technical literature review is conducted in Section 2. Section 3 describes how the final TFAs are selected and simplified. Circuit analyses and simulation results in different scenarios are provided in Section 4. Finally, Section 5 concludes the paper.

## II. Literature Review and Background

### *A. Research Method and Bibliographic Information*

This paper undertakes a comprehensive literature review of ternary adder cells till the end of 2022 to display the trajectory of this discipline. We have searched reputable publishers such as IEEE, ACM, IET, Elsevier, Springer, Wiley, Taylor & Francis, Emerald, etc for papers in which a new single-digit THA or TFA has been proposed. The Google Scholar website has also been exploited to expand our search results. “Ternary full adder” and “Ternary half adder” were the two major keywords for this purpose.

We have found 125 related papers [11-135], 20% of which have been published in IEEE journals. To be more specific, there are 15 IEEE Transactions [11-25], five IEEE Access [26-30], and two JSSC papers [31, 32]. The other two papers have been published by IEEE ELECTRON DEVICE LETT [33] and IEEE OJ-NANO [34]. Furthermore, CSSP (Springer) [35-42] and MICROELECTR J (Elsevier) [43-50] with eight papers each, and INT J ELECTRON (Taylor & Francis) [51-55] and IET CIRC DEVICE SYST (IET/IEE) [56-60] with five papers each are the journals with the highest number of publications relevant to the subject of THAs and TFAs in other publishers. IEEE ISMVL, with 11 papers [61-71], is the one that stands out among conferences. Figure 1 shows more detailed statistical information in this regard.

There is almost an equal number of THA and TFA papers in the literature (Fig. 2(a)). Additionally, 91 papers out of 125 have been published in journals (Fig. 2(b)). Finally, Fig. 2(c) shows that there has been an upward trend towards the design of THA and TFA circuits in recent years.

### *B. Statistical Information about Design Techniques*

When it comes to technical issues and implementation methods, we can categorize ternary adders in several different ways (statistical information can be found in Fig. 3):

- *Technology* (Fig. 3(a)): Early designs were based on BJT transistors. Then, being the dominant technology in the industry for decades, metal-oxide-semiconductor FET (MOSFET) has been used in 15.2% of the previous designs. However, it is the CNFET technology that with more than 65% contribution excels in popularity. This popularity is mostly because of its high flexibility in threshold adjustment. Other popular technologies are graphene nanoribbon FET (GNRFET) [34, 47, 60, 75, 76] and memristor [22, 23, 26, 63, 77-80]. A fusion of two technologies has sometimes been exploited.

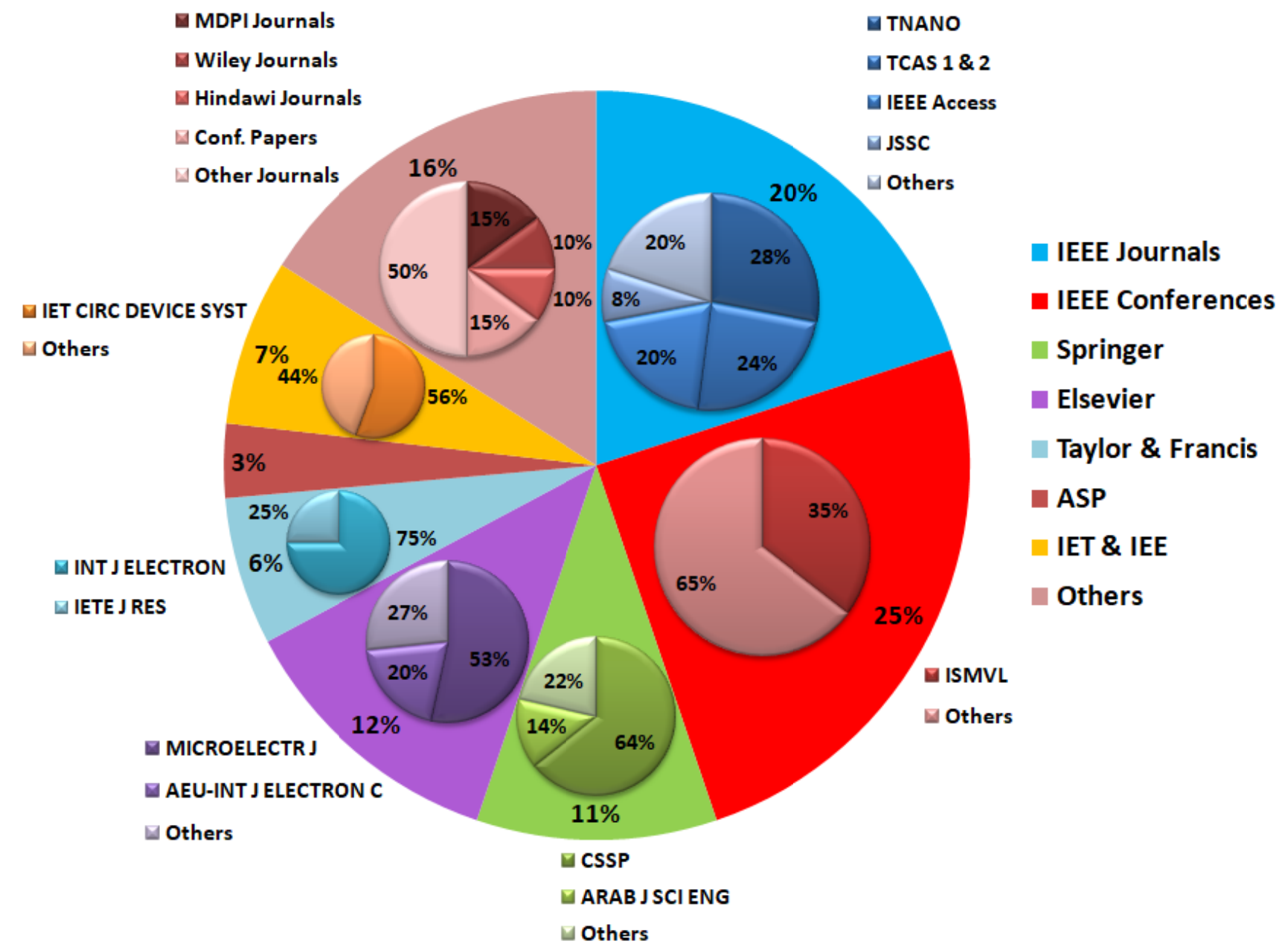

Fig. 1. Papers presenting THAs and TFAs in different publishers and journals/conferences.

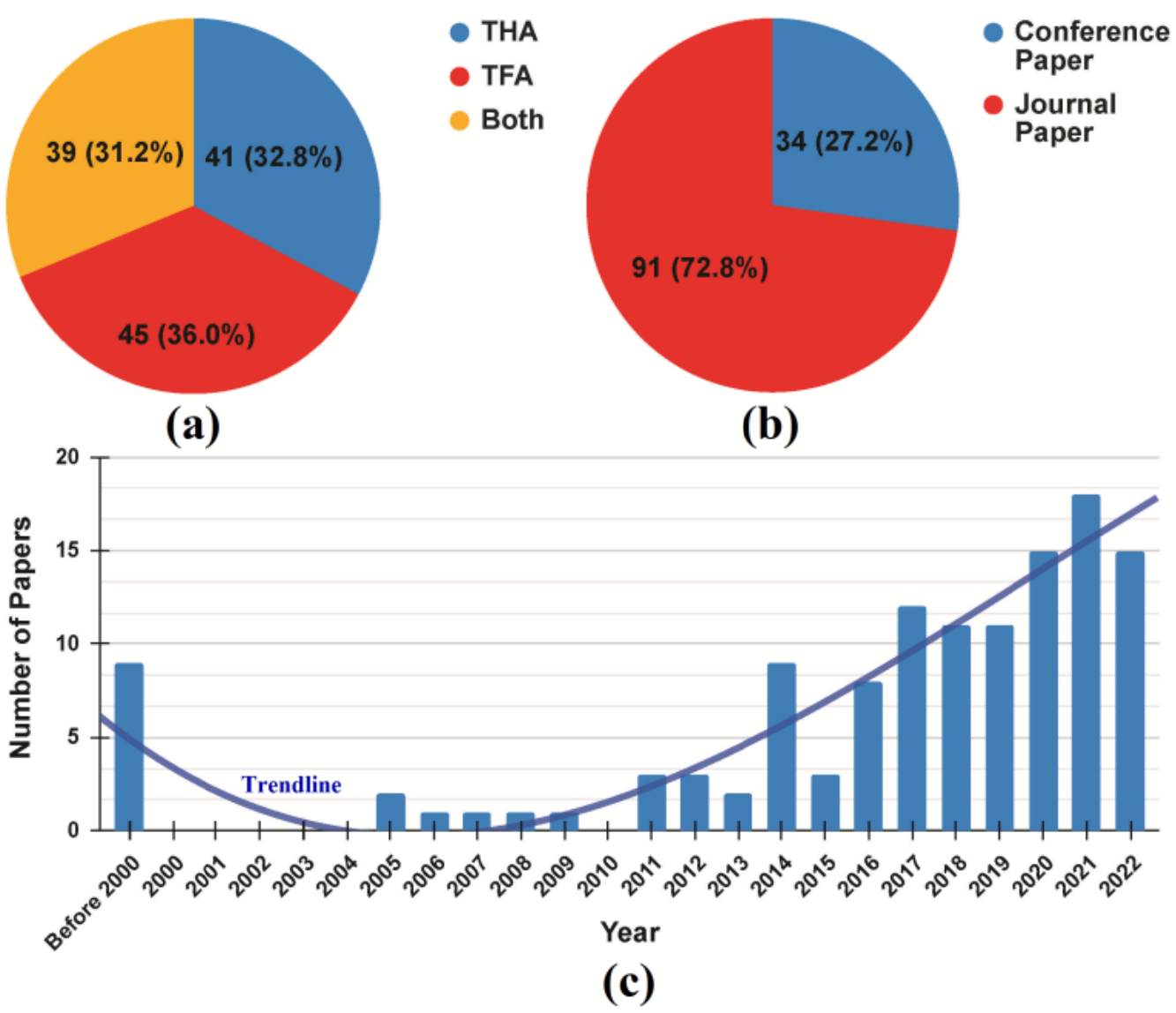

Fig. 2. Papers, (a) Proposing THA, TFA, or both, (b) Conference versus journal papers, (c) Number of papers per year.

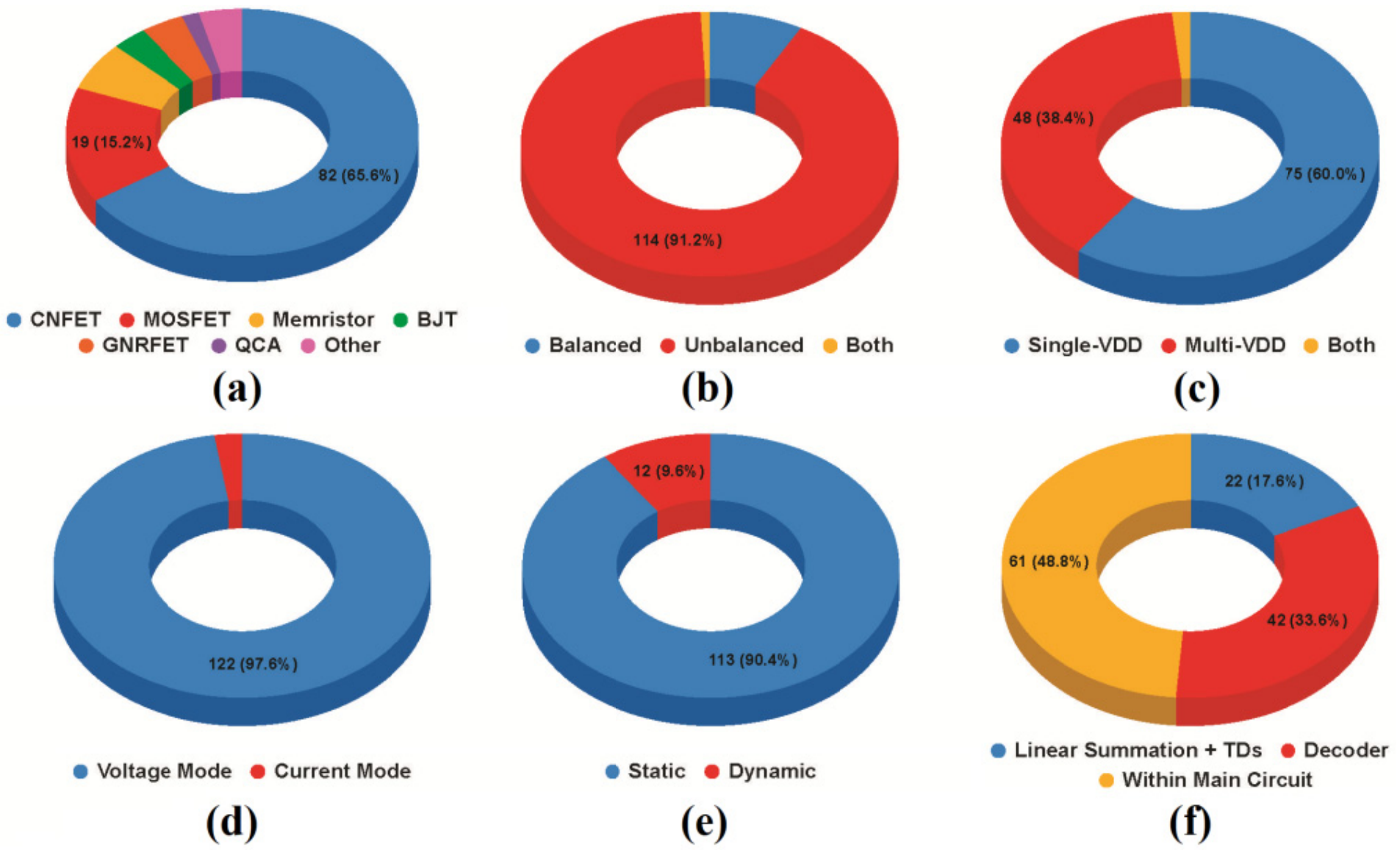

Fig. 3. Statistical information about, (a) Utilized technology, (b) Balanced versus unbalanced, (c) Single-$V_{DD}$ versus multi-$V_{DD}$, (d) Static versus dynamic, (e) Voltage mode versus current mode, (f) Input decoding methods.

- *Balanced vs. Unbalanced* (Fig. 3(b)): There are two ways to define ternary digits (trits) in base 3: *i*) Balanced (or signed) with the number set $\{-1, 0, +1\}_3$, and *ii*) Unbalanced (or unsigned) with the number set $\{0, 1, 2\}_3$. The latter is the extended version of binary logic and has been considered in more than 91% of the papers.
- *Single-$V_{DD}$ vs. Multi-$V_{DD}$* (Fig. 3(c)): The unsigned trits, 0, 1, and 2, are represented by 0V, ½$V_{DD}$, and $V_{DD}$, respectively in digital electronics. The additional voltage level, i.e. ½$V_{DD}$, can be implemented by either voltage division or an extra power supply. In a single-$V_{DD}$ scenario, the additional voltage is internally generated on a chip by using voltage division between GND and $V_{DD}$. The occurrence of voltage division is synonymous with static power dissipation. Alternatively, in a multi-$V_{DD}$ scenario, the additional voltage, which is externally generated, needs to be routed throughout the entire chip area. It imposes additional costs and wiring, increases routing complexity, and causes some parasitic effects. Following one of the main targets of MVL, which is wire reduction, 60% of the previous ternary adders are single-$V_{DD}$ designs.

- *Static vs. Dynamic* (Fig. 3(d)): Unlike static logic, where the output value only depends on the values of inputs, the output value of a dynamic circuit also depends upon a clock signal (CLK). The clock signal divides time into two phases. In the first one, the output node is pre-charged (or pre-discharged) regardless of the input values. Then, within the evaluation phase, the output node either keeps its voltage or gets discharged (charged) depending on what the values of the inputs are. This technique can help reduce transistor count. Nonetheless, a dynamic circuit requires an additional CLK signal and confronts cascading and charge sharing problems. 90.4% of the previous designs are based on static logic, and only a few are based on dynamic logic [49, 50, 61, 62, 83, 85, 88-90]. It is worth mentioning that the functionality of quantum-dot cellular automata (QCA) designs is also based on the four clock signals incorporated in this technology [82, 91].
- *Voltage Mode vs. Current Mode* (Fig. 3(e)): In voltage-mode logic (VML), logic values are represented by different voltage levels. Similarly, current levels are representative of different logic values in current-mode logic (CML). CML circuits are usually fast and can be implemented with fewer transistors. However, they dissipate much power, which is a huge downside of this logic family. There are only three CML THAs and TFAs in the literature [66, 93, 94] and the rest rely on VML.
- *Input Decoding Method* (Fig. 3(f)): There must be a way to decode input ternary signals. Some circuits, 33.6%, employ ternary decoders for this purpose. This is the most straightforward method. In a little more complex approach, instead of dealing with ternary inputs individually, the linear sum of input variables is initially obtained in some other circuits, 17.6%. To do so, the input wires are simply connected in current-mode circuits. In voltage-mode circuits, a capacitive network is required. Afterwards, threshold detectors (TDs) detect the logic level. A disadvantage associated with this method is its high sensitivity to noise and voltage variation. In MVL circuits, the entire voltage range has already been divided into several levels, and any further voltage division among the input variables causes even narrower voltage zones. On the other hand, this method can reduce the number of transistors because the linear sum of input variables is coped with cumulatively. Finally, in 48.8% of the designs, input signals are detected inside the main body of the circuit using multi-threshold transistors. Note that multi-threshold transistors are needed in every MVL circuit.

Logic or design styles define the methodologies by which logical functions can be implemented in circuit/transistor level. We classify the existing ternary circuits into 13 main logic styles, most of which are extracted from binary logic. It is also feasible to combine some of the logic styles.

1. In ternary resistor transistor logic (RTL) [11], Fig. 4(a), if both pull-down networks (PDNs) are switched off, the output node is pulled up to $V_{DD}$ through the upper resistor. If PDN1 is ON, the two resistors perform voltage division and the output becomes $\frac{1}{2}V_{DD}$. Finally, PDN2 links the output node to the ground whenever it is meant to be '0'. Compared to the resistor, PDN2 has much lower resistance and can pull the output voltage down near 0V. However, a considerable static current is dissipated in this situation. Thus, this logic style is neither full-swing nor power-efficient.
2. With almost the same structure and characteristics, the resistors in ternary RTL are replaced with constantly switched-on transistors in the ternary pseudo-NMOS logic [136], Fig. 4(b).
3. Similar to the binary CMOS logic style, the ternary counterpart (Fig. 4(c)) is composed of pull-up and pull-down networks (PUNs and PDNs), in which n-type and p-type transistors reside, respectively [12]. When PDN1 and PUN1 are both ON, the two diode-connected transistors perform voltage division, and the output voltage becomes $\frac{1}{2}V_{DD}$. In other conditions, when PDN2 and PUN2 are exclusively activated, the output value becomes '0' and '2', respectively.
4. In ternary dynamic logic [137], Fig. 4(d), the output node is initially pre-charged. Then, within the evaluation phase, it either keeps its previous value or takes a new value depending on the status of the networks. This way, one of the networks in the ternary CMOS logic style is eliminated.
5. In ternary capacitive threshold logic (CTL) [65], Fig. 4(e), the linear sum of input variables is calculated by a capacitor network. Then, the voltage level is detected by using some TDs. The rest of the circuit can be similar to the ternary CMOS logic style but with smaller networks and fewer transistors.

6. A ternary function can be represented in three different ways: *i*) Negative ternary (NT), denoted by −; *ii*) Positive ternary (PT), denoted by +; and *iii*) Standard ternary (ST). These definitions are exemplified in Table 1 for a ternary inverter (NTI, PTI, and STI). Accordingly, the ST function is the average of the other two definitions. It has been the foundation of the next logic style [56], which is called NT/PT functions in this paper (Fig. 4(f)). After *Output*− and *Output*+ are generated, two transistors divide the voltage and produce the standard output.
7. The dynamic version of the previous logic style, Fig. 4(g), for the elimination of two networks has been suggested in [90]. Two inverters are incorporated in the middle for the higher driveability of the circuit.
8. Differential cascode voltage switch (DCVS) belongs to the differential logic family where two complementary outputs are simultaneously generated. In ternary DCVSL [138], Fig. 4(h), when $O_1$ and $O_2$ are equal to $V_{DD}$ and 0V, respectively, $Out_1$ is pulled down to the ground, and $Out_2$ is connected to $V_{DD}$. The opposite conditions lead to $Out_1$=‘0’ and $Out_2$=‘2’. Finally, once $O_1$=$O_2$=$V_{DD}$, both outputs ($Out_1$ and $Out_2$) are concurrently linked to the ground and $V_{DD}$. In this case, the middle n-type and p-type transistors, which are constantly ON, perform voltage division, and both outputs become ‘1’. DCVSL circuits are usually fast, but many transistors are relatively consumed for the production of two outputs.
9. The ternary dynamic DCVSL [139], Fig. 4(i), eliminates two networks from its static version.
10. Pass transistor (PT) is a ubiquitous component for circuit design. In addition, a transmission gate (TG) is like a PT but ensures full-swing signal transmission. A network of PTs and/or TGs can be used to design a circuit (Fig. 4(j)) to have very few transistors.

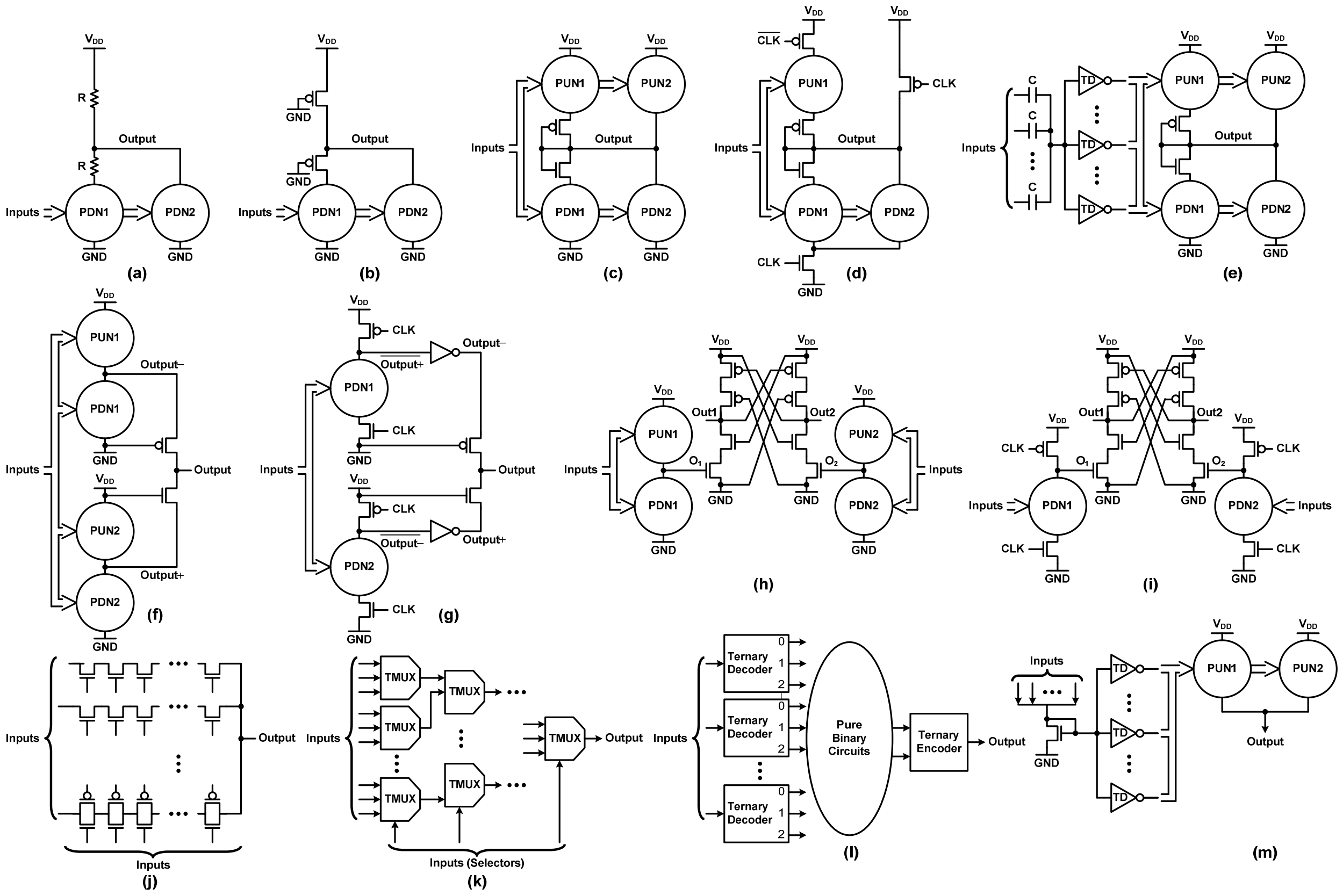

Fig. 4. Well-known logic styles for single-$V_{DD}$ unbalanced ternary circuitry, (a) Ternary RTL [11], (b) Ternary pseudo-NMOS logic [136], (c) Ternary CMOS logic style[12], (d) Ternary dynamic logic [137], (e) Ternary CTL [65], (f) NT/PT functions [56], (g) Dynamic NT/PT functions [90], (h) Ternary DCVSL [138], (i) Ternary dynamic DCVSL [139], (j) MUX-based approach [19], (k) PT/TG, (l) Decoder/Encoder [31], (m) Ternary CML [66].

TABLE I: Truth Table of NTI, PTI, and STI

| $a$ | $\bar{a}^{-}$ | $\bar{a}^{+}$ | $\bar{a}$ |
|---|---|---|---|
| 0 | 2 | 2 | 2 |
| 1 | 0 | 2 | 1 |
| 2 | 0 | 0 | 0 |

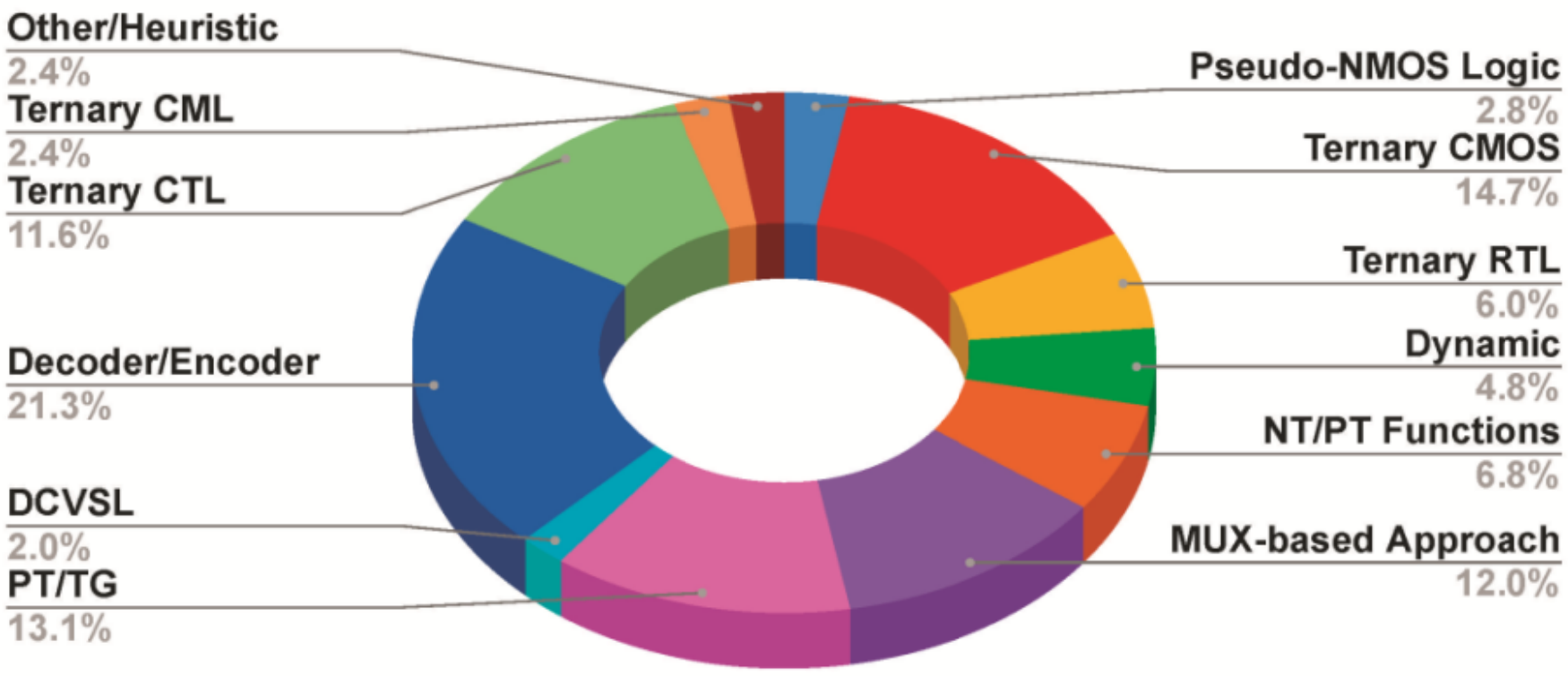

Fig. 5. Statistical information about logic styles.

11. Using multiplexers (MUXs) is one of the practical implementation methods of ternary functions. In the MUX-based approach [19], Fig. 4(k), one or more input signals are deemed to be selectors, and the rest are inputs of the multiplexer network. Multiplexers, implemented mainly by PTs/TGs, select the appropriate input value in a hierarchical structure.
12. Pure binary circuits can be used after decoding ternary input signals [31] (Fig. 4(l)). The mid-binary circuits can be in any logic style. In the end, a ternary encoder is required to produce a ternary output. There are various designs for ternary decoders and encoders in the literature.
13. In ternary current-mode logic (CML) [66], Fig. 4(m), the linear sum of input currents is initially converted to a voltage. Next, some TDs detect the voltage level. Then, a unit of current flows through the output wire if only one of the PUN1 or PUN2 is activated. When both are active, two units of current flow and the output value becomes ‘2’. Finally, there is no output current and *Output*=‘0’ if both are off. CML circuits are fast but consume power extensively.

The structures illustrated in Fig. 4 are only the basic/initial block diagrams pertinent to each logic style, and they can have several variations. Besides, in many cases, a combination of two or even more logic styles has been used. Figure 5 shows how prevalent these methods are in designing THAs and TFAs.

### *C. Complete TFA vs. Partial TFA*

The main concentration of this paper is on the design of TFAs. The truth table of a complete TFA, whose input values belong to the set $\{0, 1, 2\}_3$, is shown in Table 2. On the other hand, in a partial TFA (Table 3), the third input never becomes ‘2’ and $c \in \{0, 1\}_3$. It is feasible to show that the complete TFA is required neither for addition nor for subtraction [95]. Moreover, complete TFAs are not required for the construction of column compression ternary multipliers either [95]. Finally, division algorithms are usually based on iterative subtractions or multiplications [140]. As a result, none of the four basic arithmetic operations depends on the complete TFA.

The truth table of the partial TFA (Table 3) has nine fewer rows than the truth table of the complete TFA (Table 2). This brings an excellent opportunity to eliminate many transistors. Nevertheless, partial TFAs have been suggested in only 28.6% of the papers (Fig. 6(a)). Instead of a complete TFA, a partial one could have been presented in the other 71.4% of the previous papers. Therefore, they are not in their simplest form, and they can be simplified to a further degree. Partial TFAs are not a recent innovation, and the first one was presented in 1972 [74].

TABLE II: Truth Table of Complete Ternary Full Adder

| a | b | c = 0 | | c = 1 | | c = 2 | |
|---|---|---|---|---|---|---|---|
| | | Carry | Sum | Carry | Sum | Carry | Sum |
| 0 | 0 | 0 | 0 | 0 | 1 | 0 | 2 |
| 0 | 1 | 0 | 1 | 0 | 2 | 1 | 0 |
| 0 | 2 | 0 | 2 | 1 | 0 | 1 | 1 |
| 1 | 0 | 0 | 1 | 0 | 2 | 1 | 0 |
| 1 | 1 | 0 | 2 | 1 | 0 | 1 | 1 |
| 1 | 2 | 1 | 0 | 1 | 1 | 1 | 2 |
| 2 | 0 | 0 | 2 | 1 | 0 | 1 | 1 |
| 2 | 1 | 1 | 0 | 1 | 1 | 1 | 2 |
| 2 | 2 | 1 | 1 | 1 | 2 | 2 | 0 |

TABLE III: Truth Table of Complete Partial Full Adder

| a | b | c = 0 | | c = 1 | | c = 2 | |
|---|---|---|---|---|---|---|---|
| | | Carry | Sum | Carry | Sum | Carry | Sum |
| 0 | 0 | 0 | 0 | 0 | 1 | 0 | 2 |
| 0 | 1 | 0 | 1 | 0 | 2 | 1 | 0 |
| 0 | 2 | 0 | 2 | 1 | 0 | 1 | 1 |
| 1 | 0 | 0 | 1 | 0 | 2 | 1 | 0 |
| 1 | 1 | 0 | 2 | 1 | 0 | 1 | 1 |
| 1 | 2 | 1 | 0 | 1 | 1 | 1 | 2 |
| 2 | 0 | 0 | 2 | 1 | 0 | 1 | 1 |
| 2 | 1 | 1 | 0 | 1 | 1 | 1 | 2 |
| 2 | 2 | 1 | 1 | 1 | 2 | 2 | 0 |

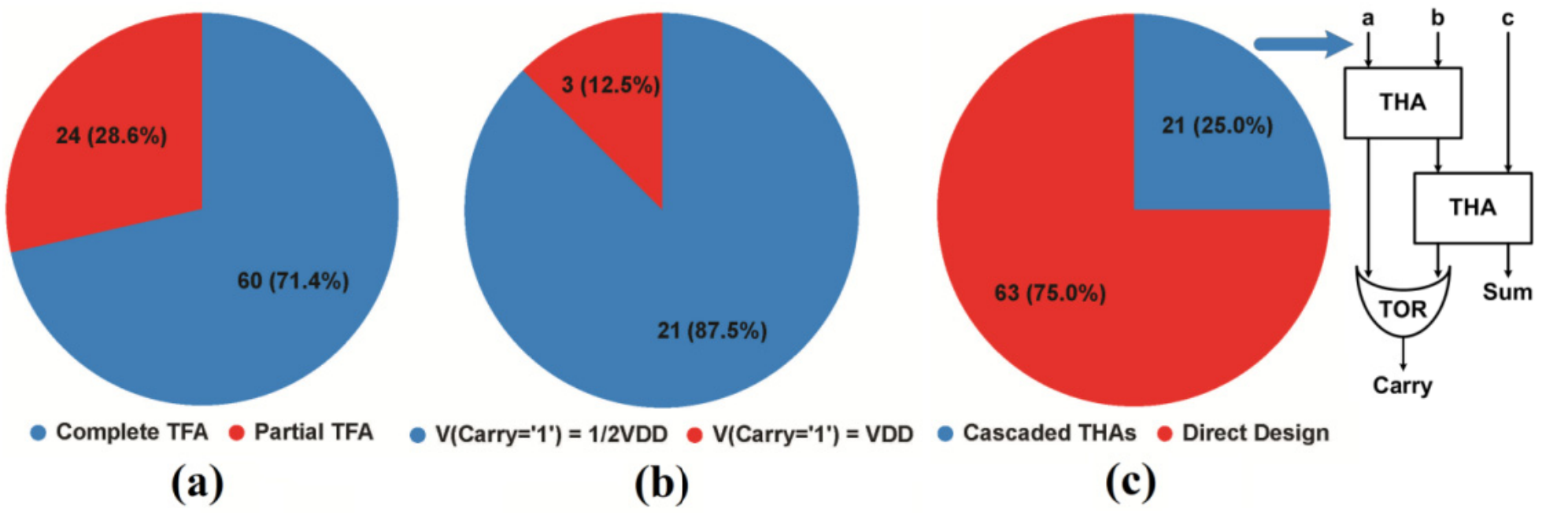

Fig. 6. Statistical information about TFAs, (a) Complete versus partial, (b) V(Carry='1') = ½$V_{DD}$ versus V(Carry='1') = $V_{DD}$, (c) Cascaded THAs versus direct design.

Additionally, the output carry has almost always been implemented by {0V, ½$V_{DD}$} in partial TFAs. In a single-$V_{DD}$ design, voltage division is inevitably required to obtain ½$V_{DD}$. Continuous static current flows every time voltage division occurs. This is the main source of high static power consumption of ternary circuits. The more voltage division happens, the more static power dissipates. Alternatively, it is possible to represent Carry='1' by $V_{DD}$ to decrease the number of times voltage division occurs in a partial TFA. In other words, instead of {0V, ½$V_{DD}$}, the number set $\{0, 1\}_3$ is implemented by {0V, $V_{DD}$} exclusively for the output carry. This way, voltage division is no longer needed for the production of the output carry signal. All we have to do is to change the design of a partial TFA in a way that the received $V_{DD}$ on its third input (input carry signal) is interpreted as logic '1'. It is worth mentioning that the way different voltage levels are interpreted is a conceptual matter. It is up to circuit designers how to map logical values onto voltage levels. Thus, the utilization of {0V, $V_{DD}$} for the output carry is completely feasible and only alters the way of interpretation for this signal. This practical technique has been employed in [96] for the first time and exploited in two other papers [37, 87]. In spite of its usefulness, this technique has been neglected in more than 87% of the previous partial TFAs (Fig. 6(b)).

Eventually, it is always possible to design a full adder by cascading two half adders. 25% of the prior TFAs are based on cascaded THAs, whereas the rest have been designed directly (Fig. 6(c)).

*D. Our Final Selection*

After the identification and taxonomic classification of the available TFAs, it is time to prune the list with the aim of reaching a limited and reasonable number of designs as a final selection. We decided to choose some TFAs whose main features are identical (e.g. all of them are single-$V_{DD}$ designs, implemented by the same technology, and representing unbalanced ternary logic) while their transistor-level layouts are different. This way, the concentration of the paper would only be on the circuit topology of TFAs. Circuit topology, especially in a single-$V_{DD}$ design where voltage division happens, directly and significantly affects performance and is a crucial factor to deal with. To reach our final selection, we decided to eliminate the following designs from the 84 existing papers in which a new design of TFA has been presented:

- Non-fully transistor-based circuits because we want to concentrate on logic styles and make our comparison independent of the utilized technology. For example, one can compare a memristor-based design with a CNFET-based design, but then the aim of comparison shifts from the topological aspect to the technological factor.
- Multi-$V_{DD}$ designs where another power rail other than $V_{DD}$ and GND is required. Again, it is possible to draw a comparison between single-$V_{DD}$ designs and multi-$V_{DD}$ ones; however, one should keep in mind that HSPICE simulator does not take any of the problematic issues caused by the extra power supply rail into account.
- Balanced ternary whose digits are not in accord with the extended binary number set. Additionally, the extra voltage of $-V_{DD}$ is often required for the implementation of balanced ternary circuits.
- CML designs which consume a lot of power.
- RTL and pseudo-NMOS circuits which are neither power-efficient nor full-swing.
- CTL designs in which the initial voltage division by the capacitor network reduces circuit robustness and makes them sensitive to noise and voltage variation.
- Dynamic logic circuits whose operations are not in accord with static circuits and depend on an external clock other than the input signals.

After eliminating the above varieties, 25 papers remain [13, 17, 18, 21, 35-37, 44, 58, 60, 67, 69, 71, 87, 95, 101-108, 131, 132]. However, many of them follow similar structures. Hence, for the sake of brevity, the next step is to remove TFAs whose structures are repetitive from our list. Our final selection from 11 papers [17, 21, 35, 37, 58, 60, 87, 95, 101-103] forms a reasonable range of TFAs with various design methodologies and logic families. The elimination of TFAs at this stage is because of their similarity to the selected ones. In addition, the same process of simplifying TFAs provided in this paper can easily be applied to any other TFA. Table 4 shows the specifications and simulation results of the 11 selected TFAs. However, as stated before, we cannot compare their delay and power values since they have been simulated in different simulation setups, including different loads, frequencies, and input patterns. Besides, there are no simulation results available in [17] and [87].

## III. Selected Ternary Full Adders

As mentioned earlier, the third input signal of a partial TFA, *c*, never becomes '2'. Therefore, this can be exploited to eliminate a multitude of transistors from a complete TFA. Table 5 shows how the simplification process can be done. For example, a p-type transistor whose threshold voltage is *Low* (*Low-$V_T$*) switches on when the input signal is '0' or '1' (*c* = '0'|'1'). Since *c* can never take any values other than '0' and '1', the transistor is always ON and can be short-circuited (replaced with a wire). Another example is when a p-type transistor with a high threshold voltage (*High-$V_T$*) is fed with negated *c*. In this situation, the transistor turns on when *c='2'*, which never happens. As a result, the transistor is always off, can be eliminated, and replaced with an *open circuit*.

Moreover, to reduce the number of times voltage division happens, we assume that the output carry voltage of a partial TFA equals $V_{DD}$ whenever it is '1' (V(Carry='1') = $V_{DD}$). As a result, the next partial TFA in a series of cascaded TFAs receives a binary signal. The way simplification is done is also exemplified in Table 5. A p-type transistor whose threshold voltage is *High* switches on when *c='0'*. In this case, the transistor must be able to distinguish between *c='0'* (=0V) and *c='1'* (=$V_{DD}$). Thus, the transistor is not removable, but it can be replaced with a *Low-$V_T$* transistor because *c* is a binary signal. A *Low-$V_T$* transistor, with a higher switching speed than a *High-$V_T$* transistor, brings about higher performance in most cases.

TABLE IV: Specifications of the Selected TFAs

| Design | Logic Style | Technology | Other Characteristics | Simulation Setup[§] | Delay[*] (ps) | Power[*] (μW) |
|---|---|---|---|---|---|---|
| [101] (2012) | Ternary CMOS | CNFET ($L_g$=32nm) | • Cascaded THAs<br>• Complete TFA | • 0.9V Power Supply<br>• 100MHz Frequency<br>• 26 Input Transitions<br>• 2fF Capacitor Load | 386.1 | 1.462 |
| [35] (2014) | Ternary CMOS | CNFET ($L_g$ N/A) | • Direct Design<br>• Complete TFA | • 0.9V Power Supply<br>• 100MHz Frequency<br>• 26 Input Transitions<br>• 2fF Capacitor Load | 166.1 | 2.209 |
| [102] (2014) | NT/PT Functions | CNFET ($L_g$=32nm) | • Cascaded THAs<br>• Complete TFA | • 0.9V Power Supply<br>• 100MHz Frequency<br>• 26 Input Transitions<br>• Ternary FO4 | 43.95 | 1.472 |
| [103] (2014) | • Sum: PT-DCVSL<br>• Carry: PT/TG | CNFET ($L_g$=32nm) | • Direct Design<br>• Partial TFA<br>• V(Carry='1') = ½$V_{DD}$ | • 0.9V Power Supply<br>• 250MHz Frequency<br>• 306 Input Transitions<br>• 2.1fF Capacitor Load | 100 | 1.45 |
| [58] (2017) | MUX-based Approach + PT/TG | CNFET ($L_g$ N/A) | • Direct Design<br>• Complete TFA | • 0.9V Power Supply<br>• Frequency N/A<br>• Transitions N/A<br>• 2fF Capacitor Load | 127.4 | 1.034 |
| [21] (2020) | NT/PT Functions | CNFET ($L_g$=32nm) | • Cascaded THAs<br>• Partial TFA<br>• V(Carry='1') = ½$V_{DD}$ | • 0.9V Power Supply<br>• 500MHz Frequency<br>• 6 Input Transitions<br>• Ternary FO4 | 269 | 0.128 |
| [60] (2020) | Decoder/Encoder + Ternary CMOS | CNFET ($L_g$=32nm) | • Direct Design<br>• Complete TFA | • 0.9V Power Supply<br>• 1GHz Frequency<br>• 26 Input Transitions<br>• No Output Load | ≈900 | ≈250 |
| [95] (2020)[†] | • 1st: CMOS<br>• 2nd: CMOS<br>• 3rd: NT/PT | CNFET ($L_g$=32nm) | • Cascaded THAs & Direct Design<br>• Partial TFA<br>• V(Carry='1') = ½$V_{DD}$ | • 0.9V Power Supply<br>• 100MHz Frequency<br>• 26 Input Transitions<br>• Ternary FO4 | 108.37<br>66.168<br>40.966 | 1.8916<br>1.9341<br>1.4096 |
| [87] (2021) | MUX-based Approach + PT/TG + Ternary CMOS | CNFET ($L_g$ N/A) | • Direct Design<br>• Partial TFA<br>• V(Carry='1') = $V_{DD}$ | Not Available (N/A) | | |
| [17] (2021) | NT/PT Functions | CNFET ($L_g$=45nm) | • Cascaded THAs<br>• Complete TFA | • 1V Power Supply<br>• 100MHz Frequency<br>• 26 Input Transitions<br>• 1fF Capacitor Load | No Simulation Results for the Proposed TFA | |
| [37] (2021) | Ternary CMOS + PT/TG | CNFET ($L_g$=32nm) | • Direct Design<br>• Partial TFA<br>• V(Carry='1') = $V_{DD}$ | • 0.9V Power Supply<br>• 250MHz Frequency<br>• 64 Input Transitions<br>• 2fF Capacitor Load | 238.94 | 0.98 |

[*] Simulation results reported in the original papers.
[§] All of the simulations have been performed at room temperature with no input buffers.
[†] Three different designs have been presented in this paper.

TABLE V: Simplification and Transistor Elimination Process

| Transistor in Complete TFA | Functionality Description | Equivalent Component in Partial TFA |
|---|---|---|
| **P-Type Transistor** | | |
| 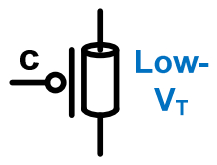 | Switches on when c = ‘0’\|‘1’ => Always ON | Wire |
| 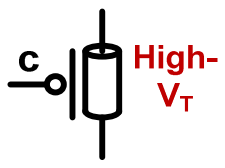 | Switches on when c = ‘0’ | 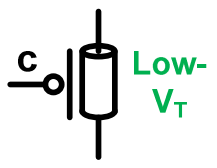 |
| 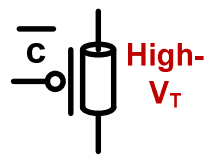 | Switches on when c = ‘2’ => Always Off | Open Circuit |
| 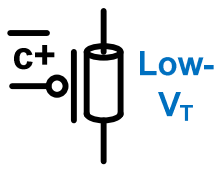 | Switches on when c = ‘2’ => Always Off | Open Circuit |
| 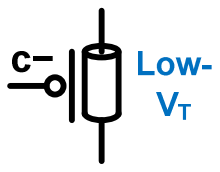 | Switches on when c = ‘0’\|‘1’ => Always ON | Wire |
| **N-Type Transistor** | | |
| 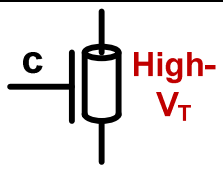 | Switches on when c = ‘2’ => Always Off | Open Circuit |
| 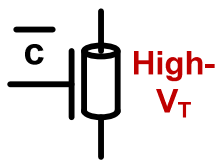 | Switches on when c = ‘0’ | 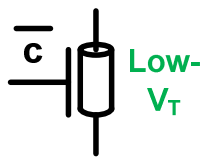 |
| 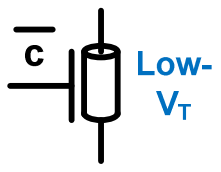 | Switches on when c = ‘0’\|‘1’ => Always ON | Wire |
| 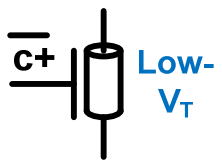 | Switches on when c = ‘0’\|‘1’ => Always ON | Wire |
| 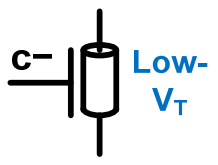 | Switches on when c = ‘2’ => Always Off | Open Circuit |

*A. Designs in [101] and [95]*

The complete TFA in [101] (Fig. 7), which is based on ternary CMOS, has been transformed into a partial TFA in [95], where V(Carry=‘1’) = ½$V_{DD}$. In this paper, we propose another version where V(Carry=‘1’) = $V_{DD}$ (Fig. 8). The STI that produces negated *c* is replaced with a binary inverter (the added transistors are indicated by green color). Moreover, some n-type transistors are crossed out since the voltage division is no longer required in the *Carry* generator part. They were responsible for making V(Carry) ½$V_{DD}$. Table 6 shows how efficient the new partial TFA compared to the previous versions is. It has 32 fewer transistors and a 70.2% lower power-delay product (PDP) than the original design [101]. PDP, calculated by (1), is an important evaluating factor in VLSI design, which makes a balance between delay and power. In addition, because no voltage division occurs in the *Carry* generator part, it consumes 18.2% less power than the one in [95].

$$PDP = Avg(Power) \times Max(Delay) \quad (1)$$

TABLE VI: Simulation Results for the TFA in [101] and Its Simplified Versions

| TFA | Delay (ps) | Power (μW) | PDP (fJ) | #Transistor |
|---|---|---|---|---|
| Complete [101] | 180.70 | 7.8403 | 1.4167 | 106 |
| Partial [95] | 125.26 | 7.6817 | 0.9622 | 87 |
| This Paper | 67.210 | 6.2825 | 0.4223 | 74 |
| Improvement w.r.t. [101] | ≈62.8% | ≈19.9% | ≈70.2% | ≈30.2% |

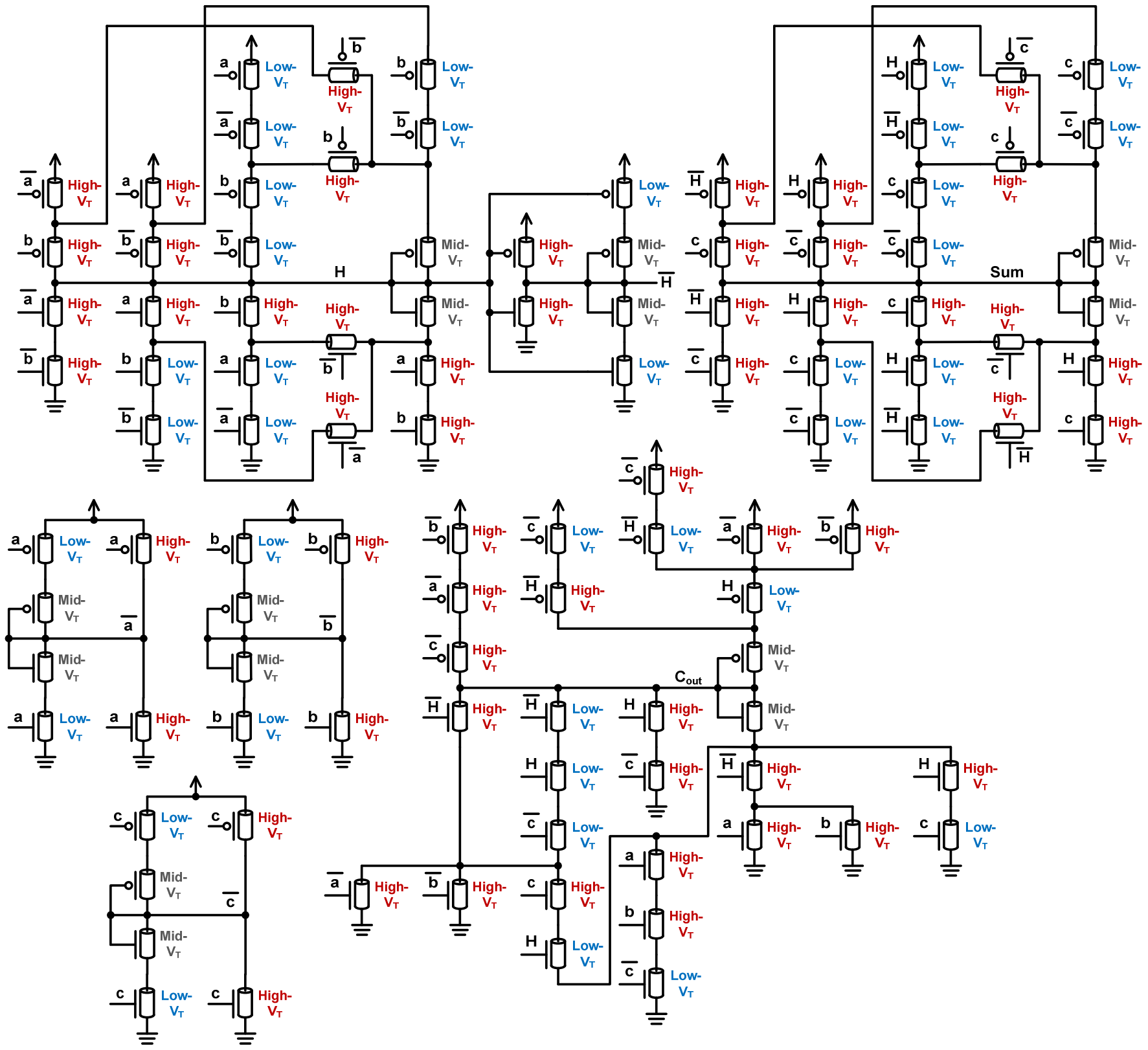

Fig. 7. The complete TFA presented in [101].

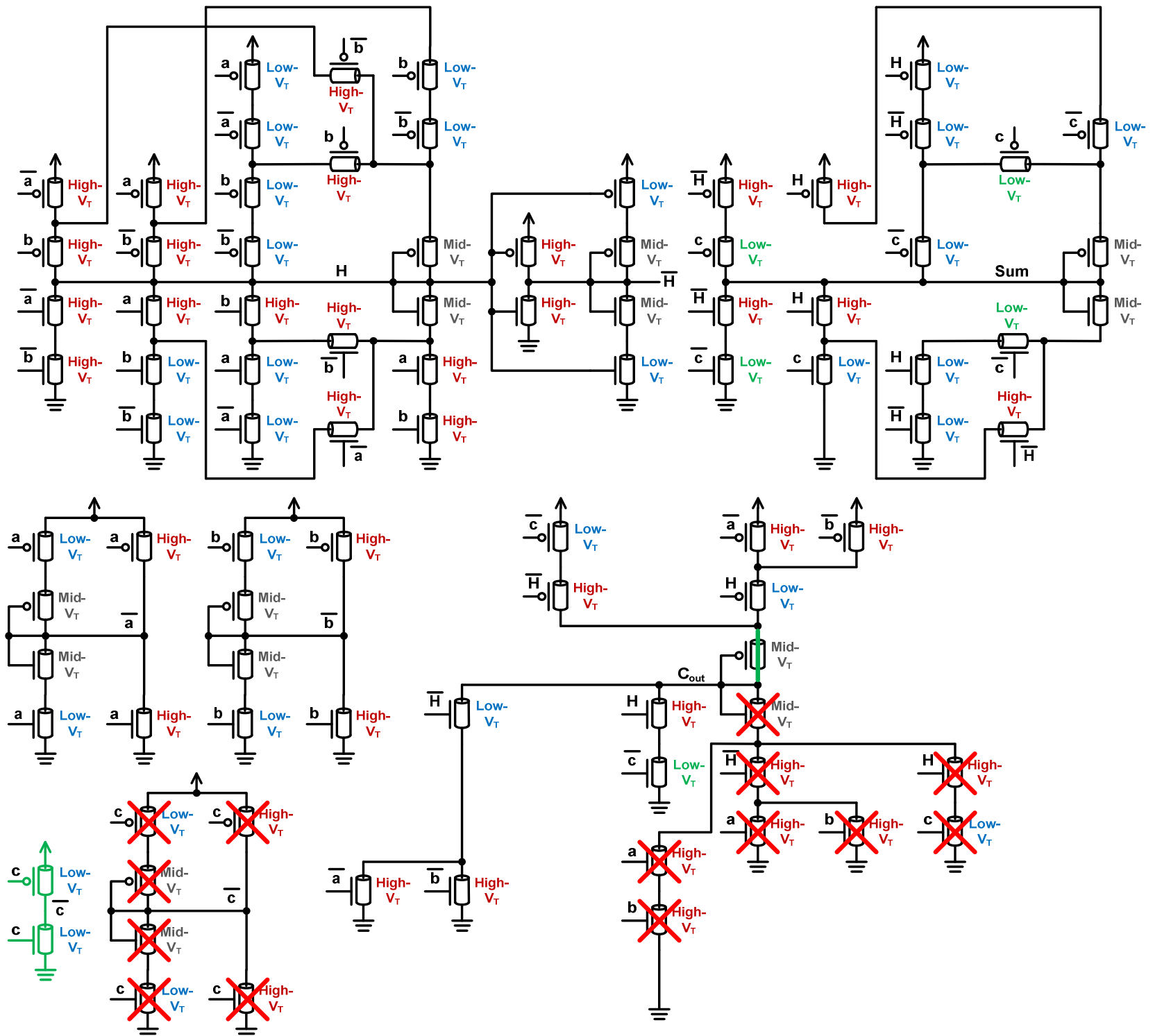

Fig. 8. The partial TFA in [95] is simplified in this paper to reach a new partial TFA, where V(Carry=‘1’) = $V_{DD}$.

*B. Designs in [35] and [95]*

Another complete TFA, based on ternary CMOS, has been presented in [35] (Fig. 9). It has been changed into a partial TFA in [95]. We simplify the partial TFA even further by considering V(Carry='1') to be $V_{DD}$ (Fig. 10). The whole simplification process is precisely similar to what was explained for the previous adder cell. Table 7 shows the simulation results for the prior and new designs. The new partial TFA has 56 fewer transistors than the original one [35] and improves its PDP by 46.3%.

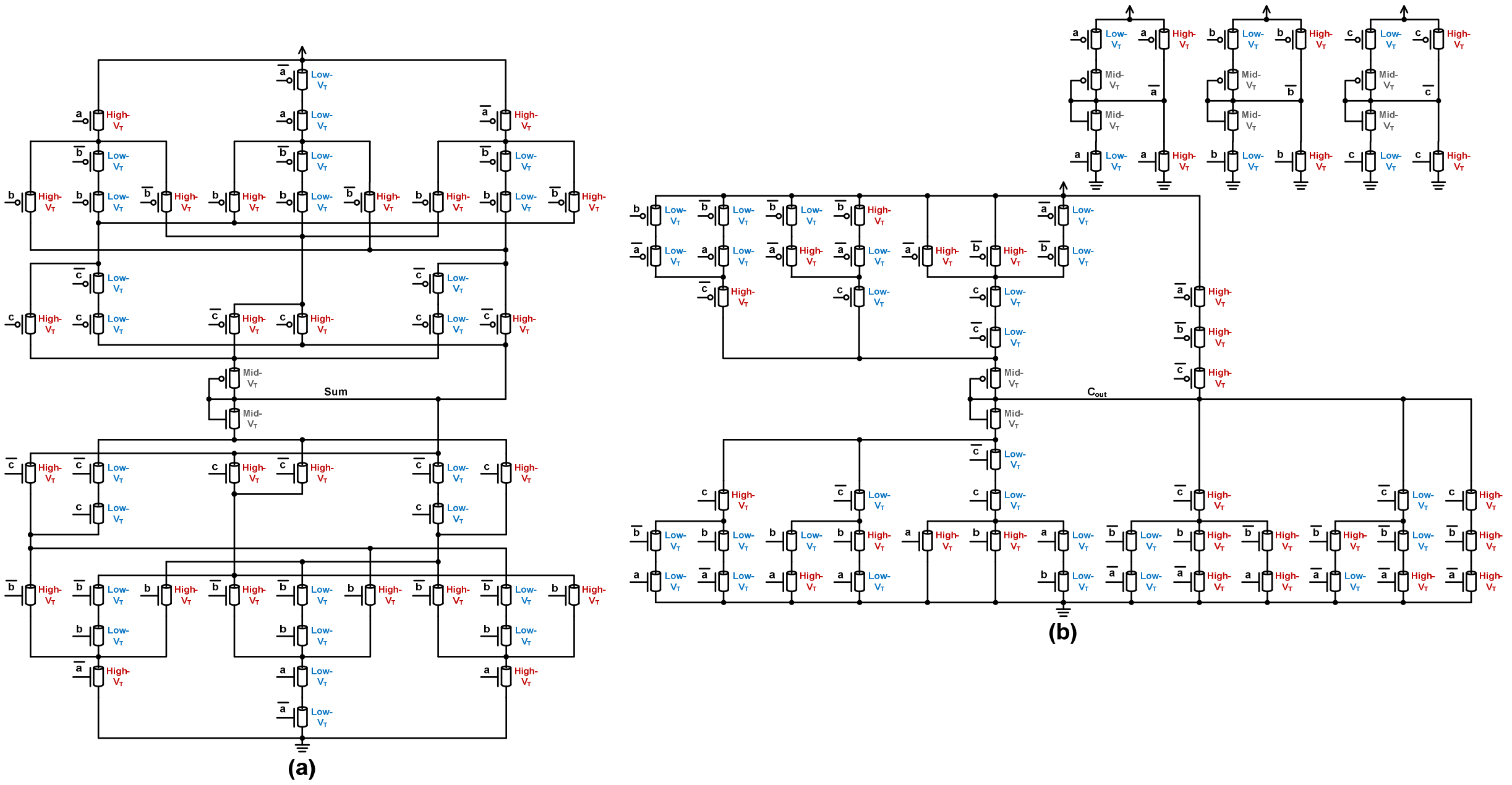

Fig. 9. The complete TFA presented in [35].

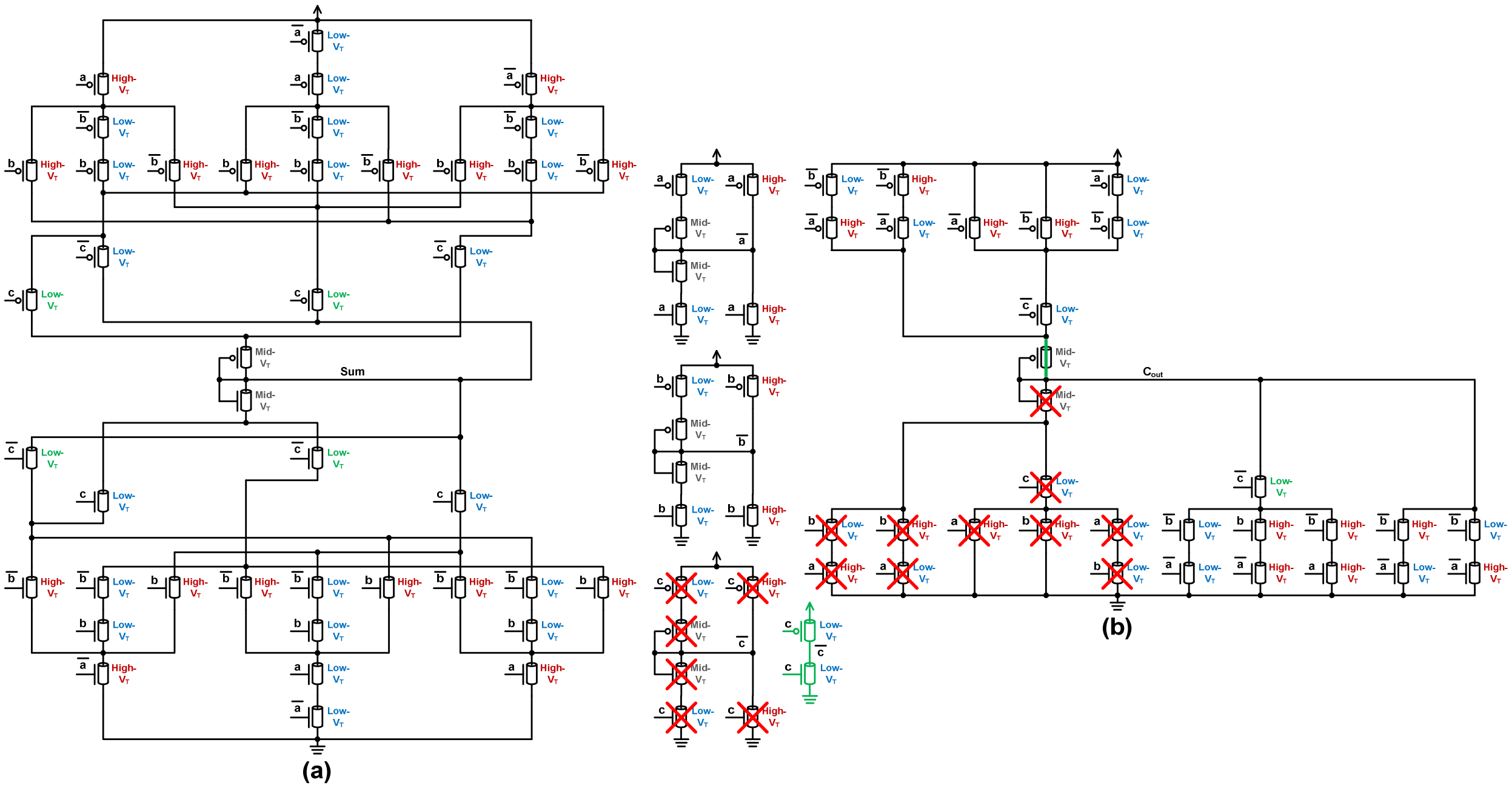

Fig. 10. The partial TFA in [95] is simplified in this paper to reach a new partial TFA, where V(Carry='1') = $V_{DD}$.

TABLE VII: Simulation Results for the TFA in [35] and Its Simplified Versions

| TFA | Delay (ps) | Power (μW) | PDP (fJ) | #Transistor |
|---|---|---|---|---|
| Complete [35] | 119.14 | 8.0044 | 0.9405 | 132 |
| Partial [95] | 117.52 | 7.3571 | 0.8646 | 91 |
| This Paper | 111.94 | 4.5080 | 0.5046 | 76 |
| Improvement w.r.t. [35] | ≈6.04% | ≈43.7% | ≈46.3% | ≈42.4% |

### *C. Designs in [102] and [95]*

A complete TFA, which is based on NT/PT functions (Fig. 11), and its partial counterpart have been presented in [102] and [95], respectively. In our proposed TFA (Fig. 12), one (or two) of the duplicated transistors in parallel paths (highlighted by different colors) is/are factored out and eliminated. Simulation results can be seen in Table 8.

TABLE VIII: Simulation Results for the TFA in [102] and Its Simplified Versions

| TFA | Delay (ps) | Power (μW) | PDP (fJ) | #Transistor |
|---|---|---|---|---|
| Complete [102] | 85.024 | 6.2898 | 0.5348 | 142 |
| Partial [95] | 66.881 | 5.8943 | 0.3942 | 112 |
| This Paper | 58.191 | 5.3980 | 0.3141 | 102 |
| Improvement w.r.t. [102] | ≈31.6% | ≈14.2% | ≈41.3% | ≈28.2% |

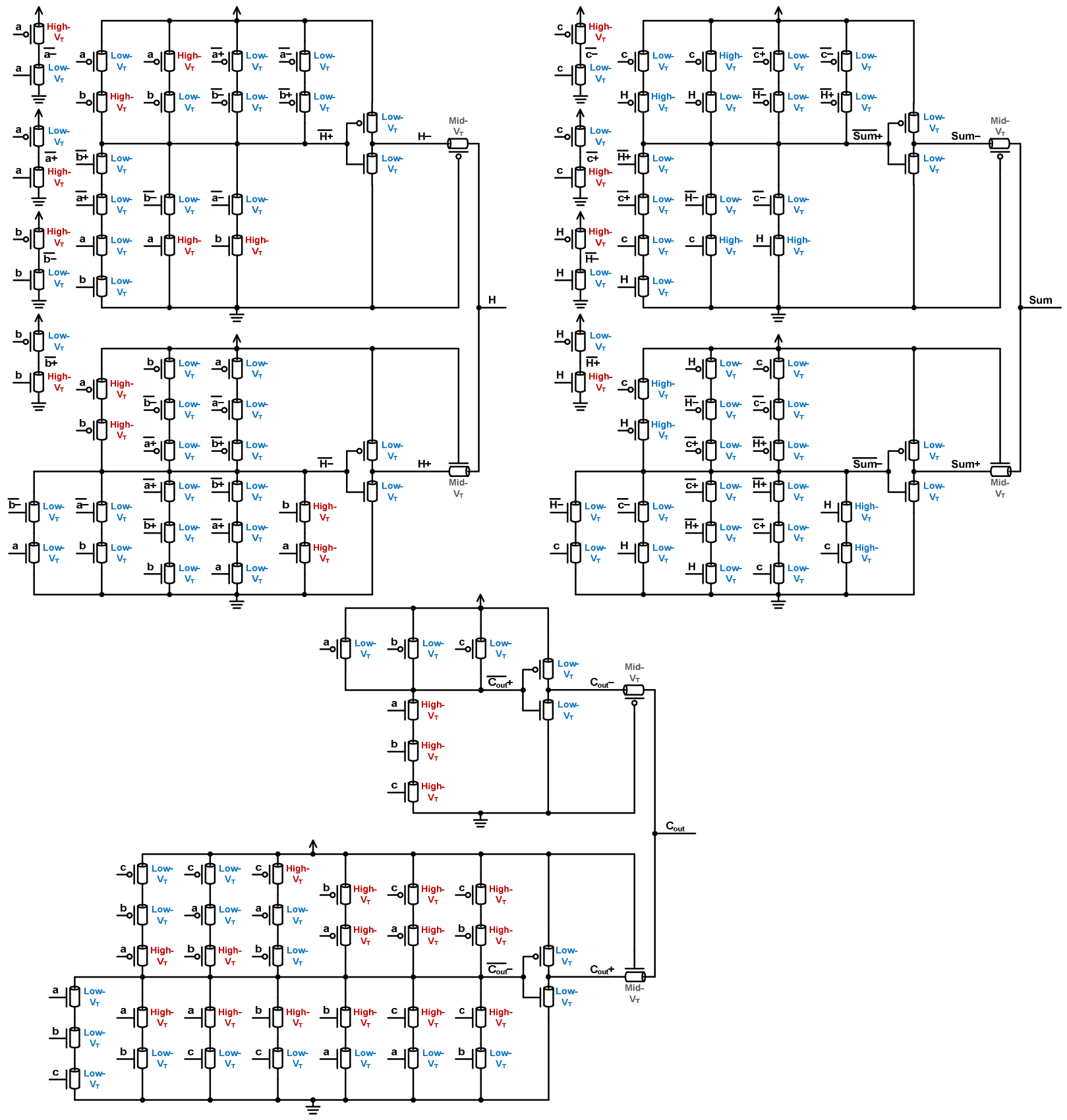

Fig. 11. The complete TFA presented in [102].

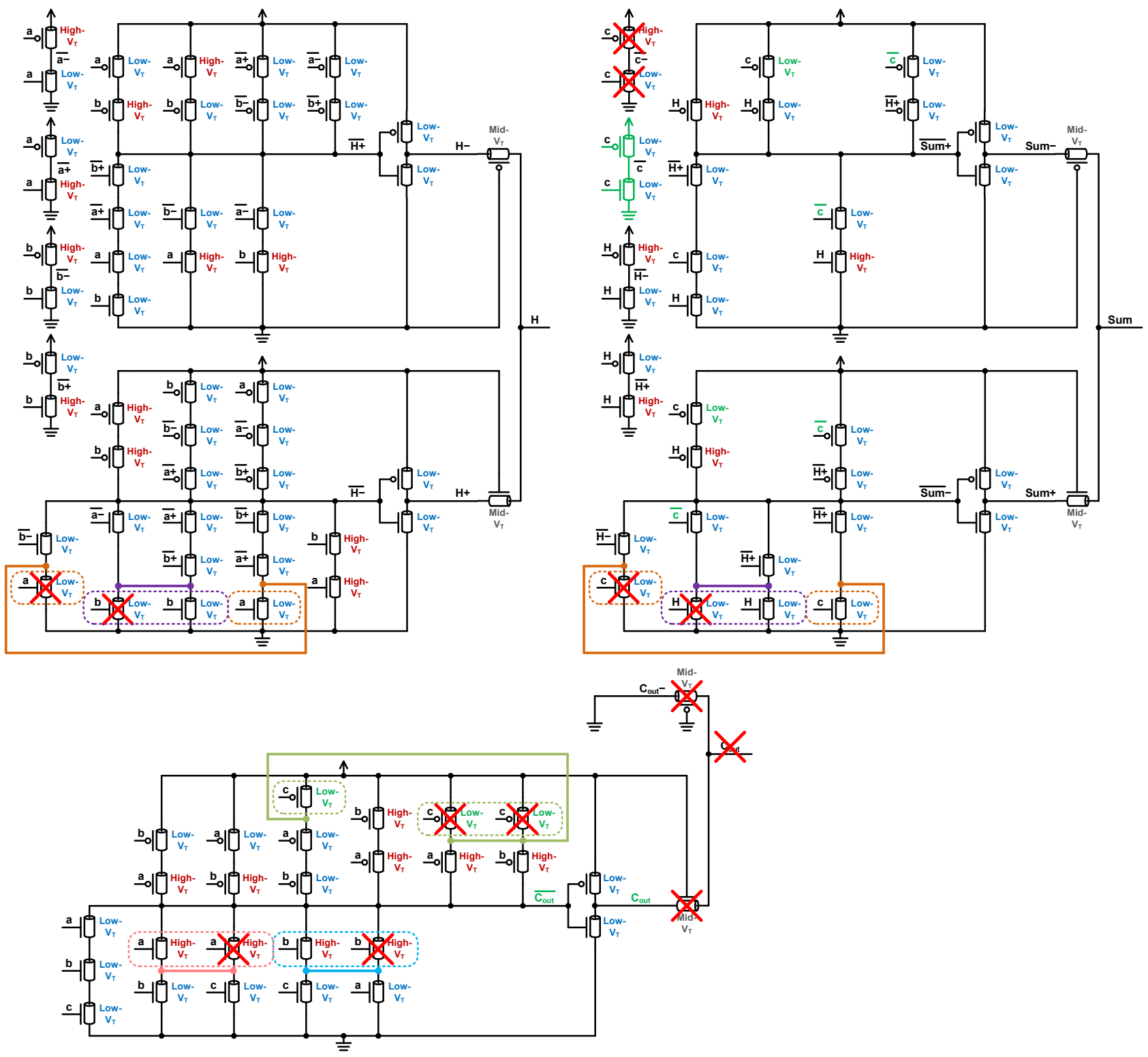

Fig. 12. The partial TFA in [95] is simplified in this paper to reach a new partial TFA, where V(Carry='1') = $V_{DD}$.

*D. Design in [103]*

The original ternary adder cell presented in [103] is a partial TFA in which V(Carry='1') = ½$V_{DD}$. The first major drawback of this design is that the voltage level of ½$V_{DD}$ is generated internally by a non-stop voltage division. It is the main source of enormous power dissipation. The second disadvantage is the utilization of n-type PTs that do not pass $V_{DD}$ in a full-swing manner. In our paper, measures are taken to improve its performance (Fig. 13). First, the PT network for the generation of the output carry is fed with $V_{DD}$, and the corresponding PTs are changed to p-type. This way, $V_{DD}$ is also passed through in a full-swing manner. The middle PTs are changed to TGs as well. Second, the low-$V_T$ n-type PTs in the *Sum* generator part are replaced with ultra-low-$V_T$ ones to alleviate the problem of voltage drop. Table 9 shows how much these modifications improve the performance, although the number of transistors remains unchanged.

TABLE IX: Simulation Results for the TFA in [103] and Its Simplified Version

| TFA | Delay (ps) | Power (µW) | PDP (fJ) | #Transistor |
|---|---|---|---|---|
| Complete [103] | 62.416 | 9.7617 | 0.6093 | 100 |
| This Paper | 56.936 | 5.3516 | 0.3047 | 100 |
| Improvement | ≈8.78% | ≈45.2% | ≈50.0% | 0% |

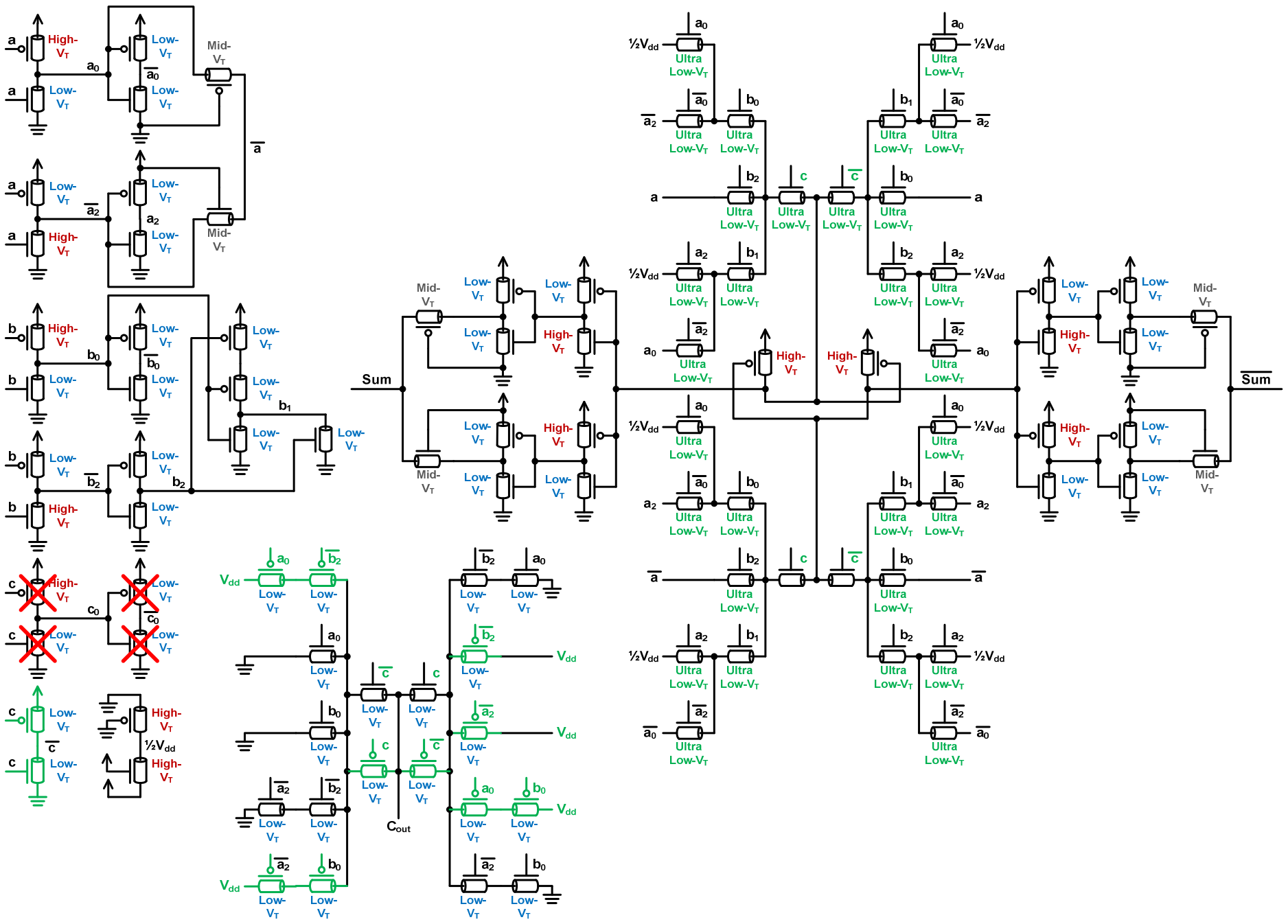

Fig. 13. The partial TFA in [103] is simplified in this paper to reach a new partial TFA, where V(Carry=‘1’) = $V_{DD}$.

### *E. Design in [58]*

The complete TFA in [58] is almost entirely based on TGs. After some alterations, the new partial TFA (Fig. 14) has 43 fewer transistors and eliminates the continuous voltage division for the production of ½$V_{DD}$. By considering V(Carry=‘1’) = $V_{DD}$, the internal generation of ½$V_{DD}$ is not needed anymore. Besides, instead of a TG, a PT can be deployed when a fixed voltage is supposed to be passed. The simulation results for both versions are presented in Table 10.

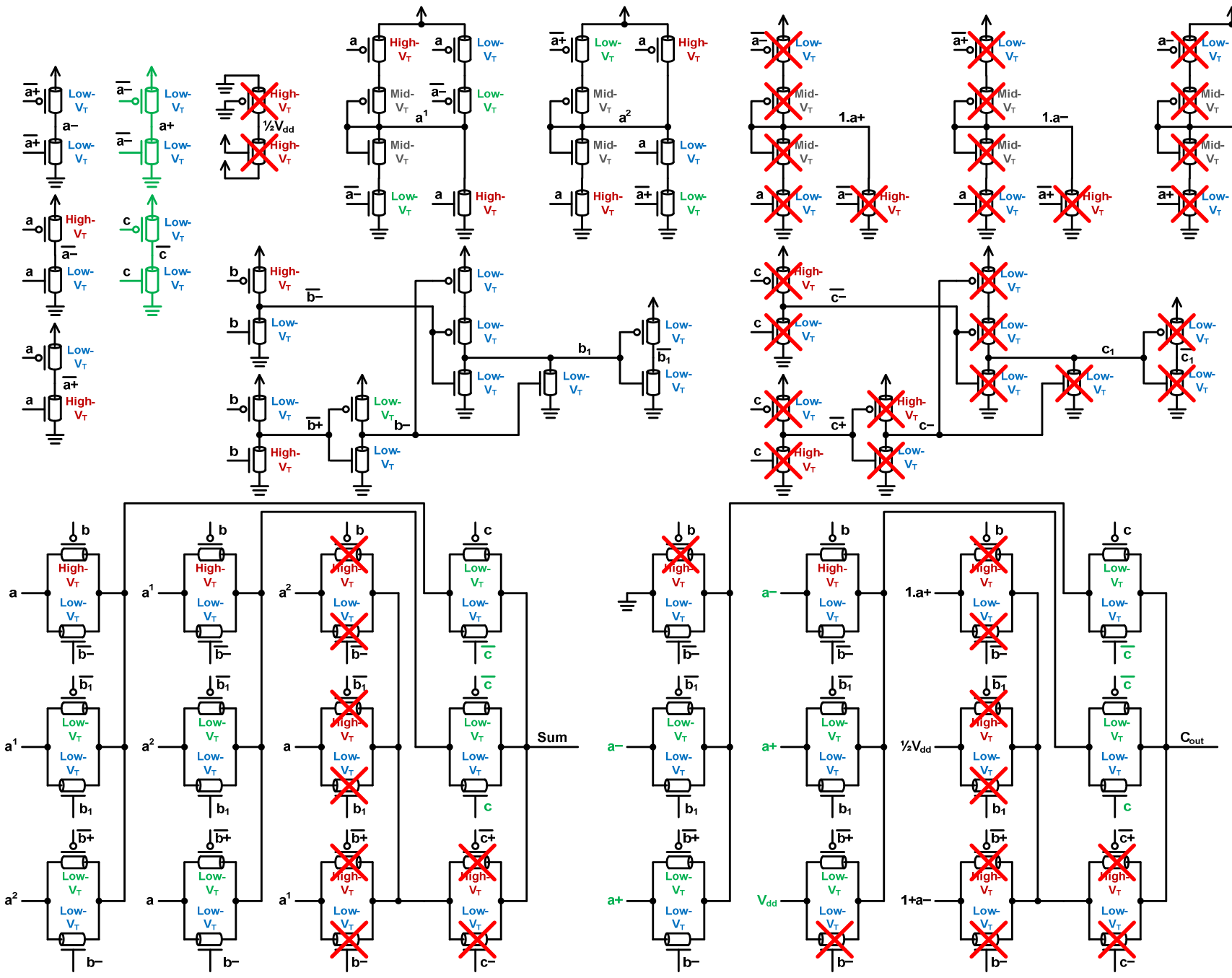

Fig. 14. The complete TFA in [58] is simplified in this paper to reach a partial TFA, where V(Carry=‘1’) = $V_{DD}$.

TABLE X: Simulation Results for the TFA in [58] and Its Simplified Version

| TFA | Delay (ps) | Power (μW) | PDP (fJ) | #Transistor |
|---|---|---|---|---|
| Complete [58] | 116.79 | 10.338 | 1.2074 | 109 |
| This Paper | 90.231 | 6.1885 | 0.5584 | 66 |
| Improvement | ≈22.7% | ≈40.1% | ≈64.9% | ≈39.4% |

*F. Design in [21]*

The partial TFA in [21] is another design based on NT/PT functions. In our simplified version, since the output carry is deemed either 0V or $V_{DD}$, three voltage divisions in the *Carry* generator part are eliminated (Fig. 15). As a result of this simplification, the average power consumption decreases by 26.9% (Table 11).

TABLE XI: Simulation Results for the TFA in [21] and Its Simplified Version

| TFA | Delay (ps) | Power (μW) | PDP (fJ) | #Transistor |
|---|---|---|---|---|
| Complete [21] | 227.82 | 6.1240 | 1.3952 | 106 |
| This Paper | 154.63 | 4.4792 | 0.6926 | 78 |
| Improvement | ≈32.1% | ≈26.9% | ≈50.4% | ≈26.4% |

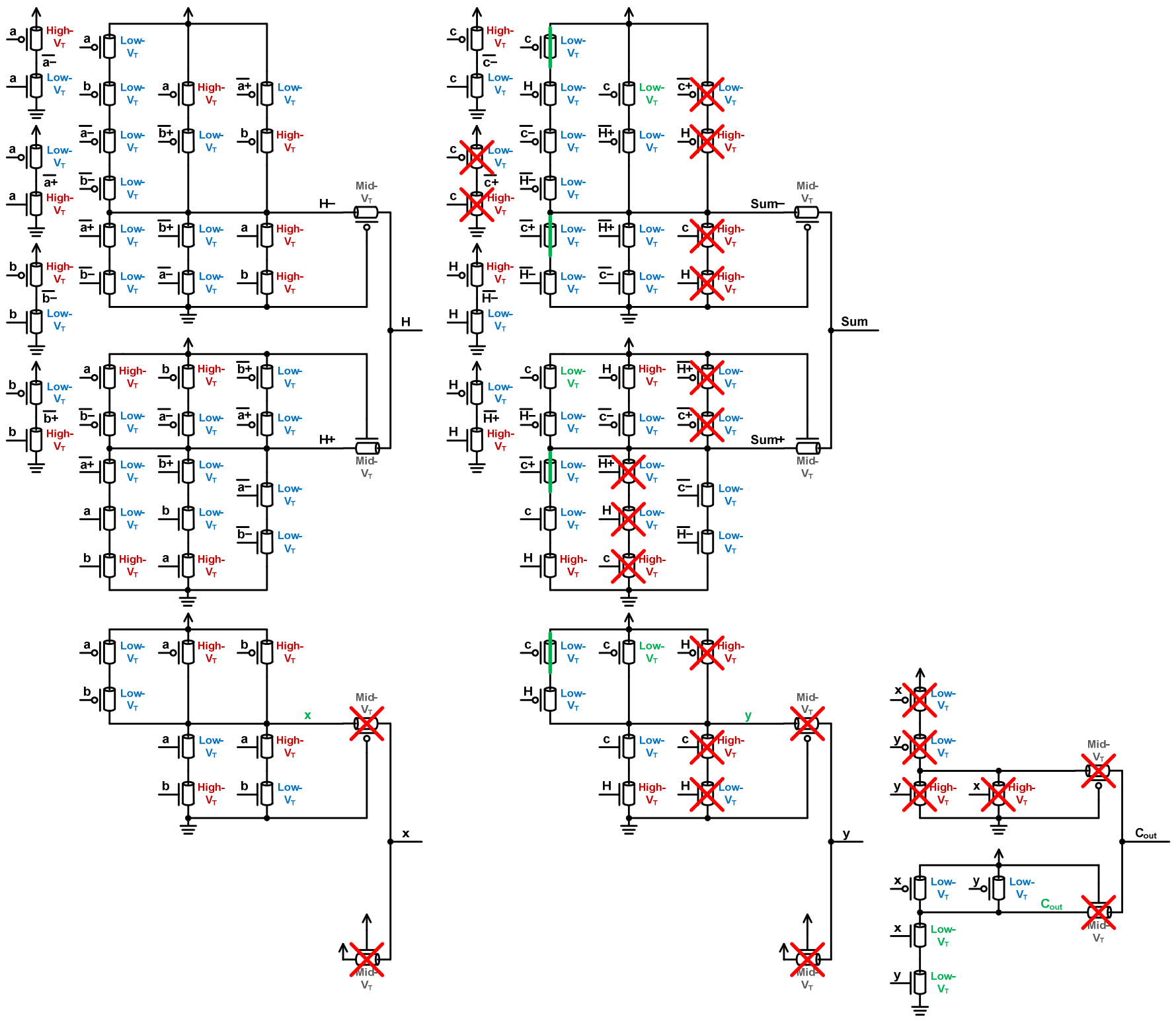

Fig. 15. The partial TFA in [21] is simplified in this paper to reach a new partial TFA, where V(Carry='1') = $V_{DD}$.

*G. Design in [60]*

The complete TFA in [60] is an example where the ternary inputs are initially decoded. The decoded signals are binary. However, ternary logic gates have still been used for the main body of the circuit. This is why it has too many transistors (1130 ones), which is not acceptable at all. The authors could have replaced ternary logic gates with binary counterparts. By considering this replacement, changing the whole adder cell into a partial one, and eliminating repetitive AND gates, 864 transistors are deleted in our simplified version (Fig. 16). The simulation results are depicted in Table 12. The original TFA fails to operate at 1GHz input frequency because of its long delay.

TABLE XII: Simulation Results for the TFA in [60] and Its Simplified Version

| TFA | Delay (ps) | Power (μW) | PDP (fJ) | #Transistor |
|---|---|---|---|---|
| Complete [60] | Failed | Failed | Failed | 1130 |
| This Paper | 318.49 | 7.8805 | 2.5098 | 266 |
| Improvement | - | - | - | ≈76.5% |

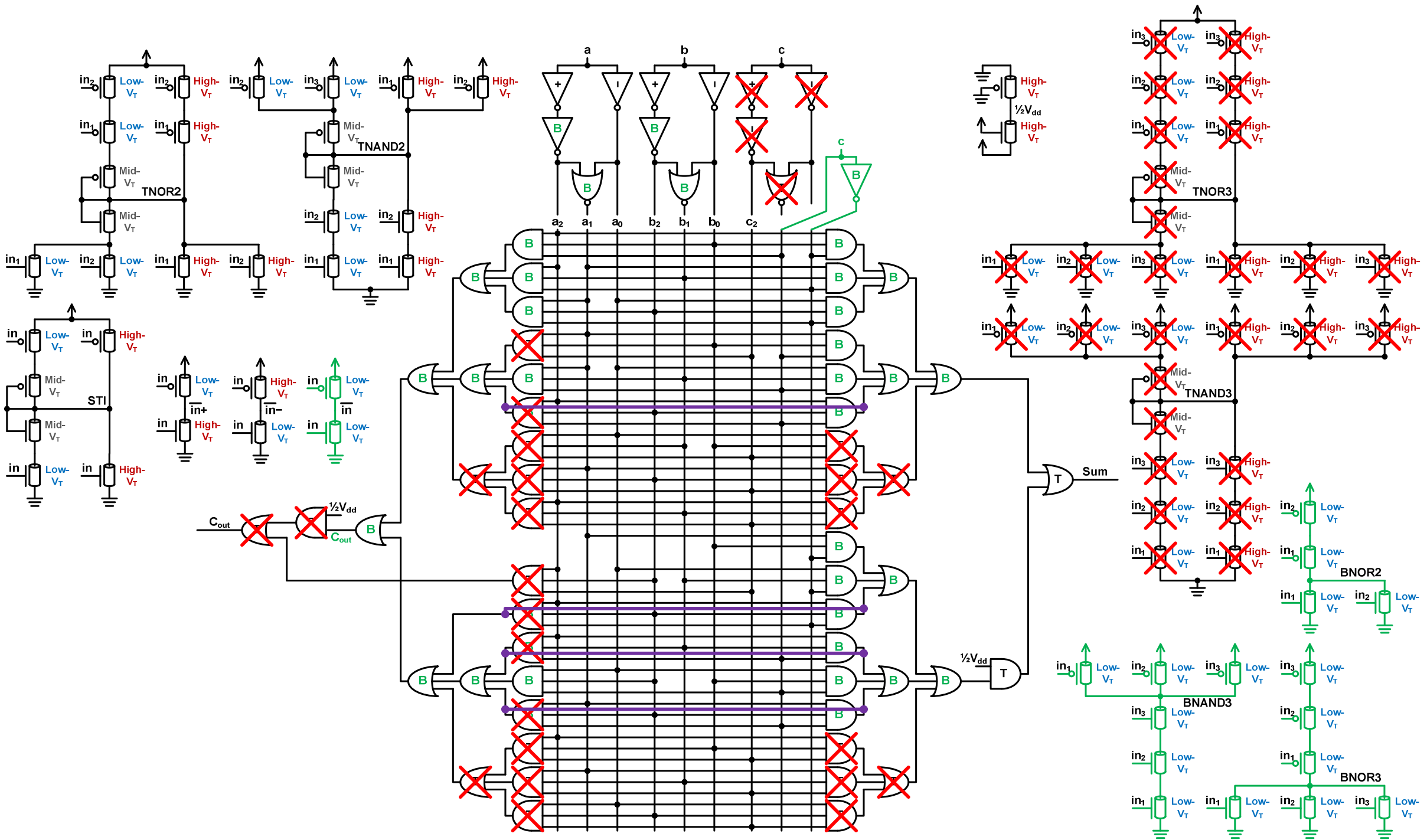

Fig. 16. The complete TFA in [60] is simplified in this paper to reach a partial TFA, where V(Carry='1') = $V_{DD}$.

### H. Design in [87]

The partial TFA in [87], where V(Carry='1') = $V_{DD}$, is in its ideal form. The only improvement possible to make is to remove four transistors from the original design (Fig. 17). Table 13 shows the simulation results for both versions.

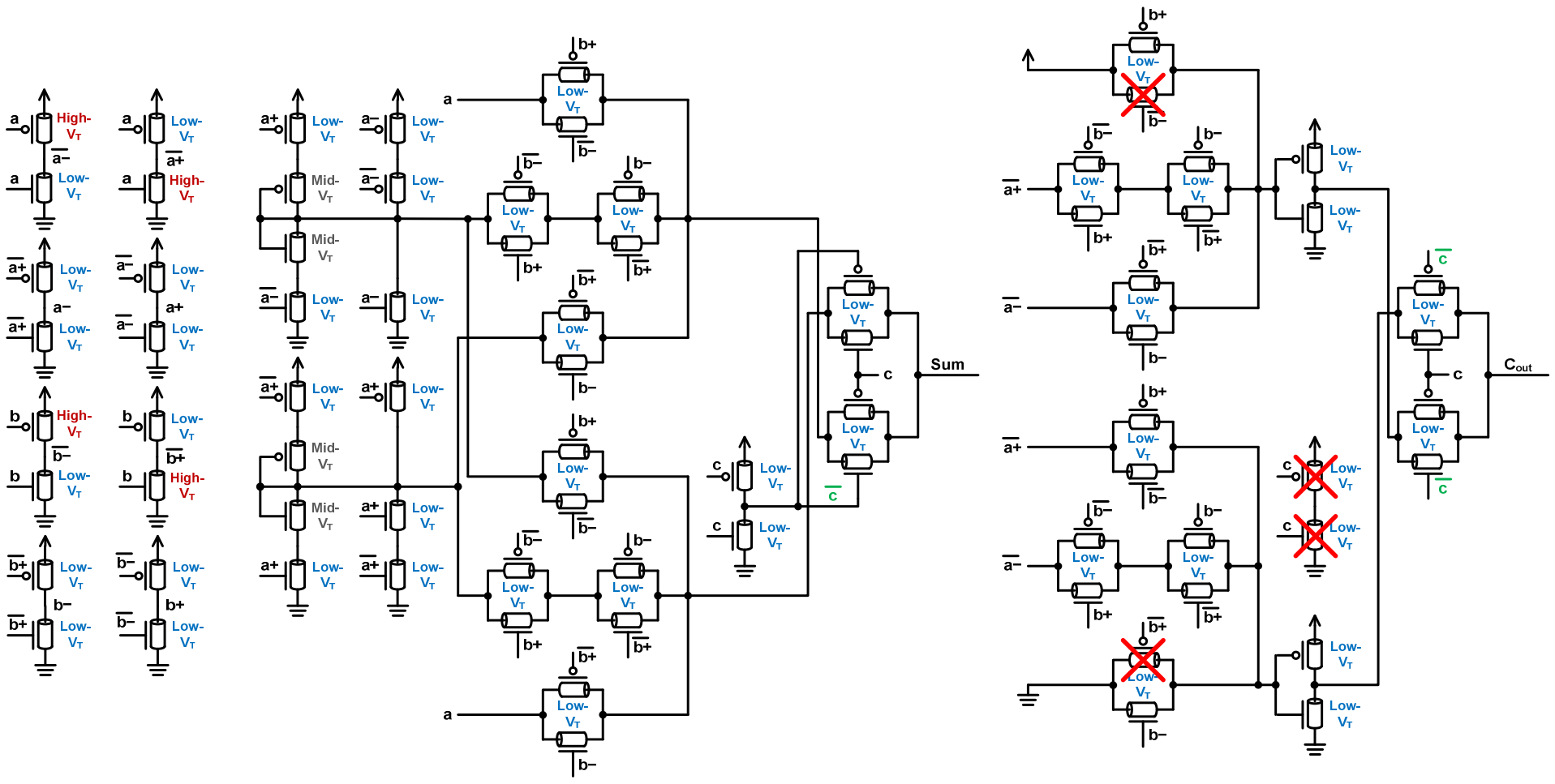

Fig. 17. The partial TFA in [87] is simplified in this paper.

TABLE XIII: Simulation Results for the TFA in [87] and Its Simplified Version

| TFA | Delay (ps) | Power (μW) | PDP (fJ) | #Transistor |
|---|---|---|---|---|
| Complete [87] | 67.556 | 5.5509 | 0.3750 | 78 |
| This Paper | 66.665 | 5.4544 | 0.3636 | 74 |
| Improvement | ≈1.32% | ≈1.74% | ≈3.04% | ≈5.13% |

*I. Design in [17]*

The complete TFA in [17] is simplified in this paper by the elimination of 84 transistors (Fig. 18). Table 14 shows that the simplified version is much faster (24.8%) and more power efficient (22.8%) than the previous version.

TABLE XIV: Simulation Results for the TFA in [17] and Its Simplified Version

| TFA | Delay (ps) | Power (μW) | PDP (fJ) | #Transistor |
|---|---|---|---|---|
| Complete [17] | 91.203 | 9.2611 | 0.8446 | 198 |
| This Paper | 68.622 | 7.1485 | 0.4905 | 114 |
| Improvement | ≈24.8% | ≈22.8% | ≈41.9% | ≈42.4% |

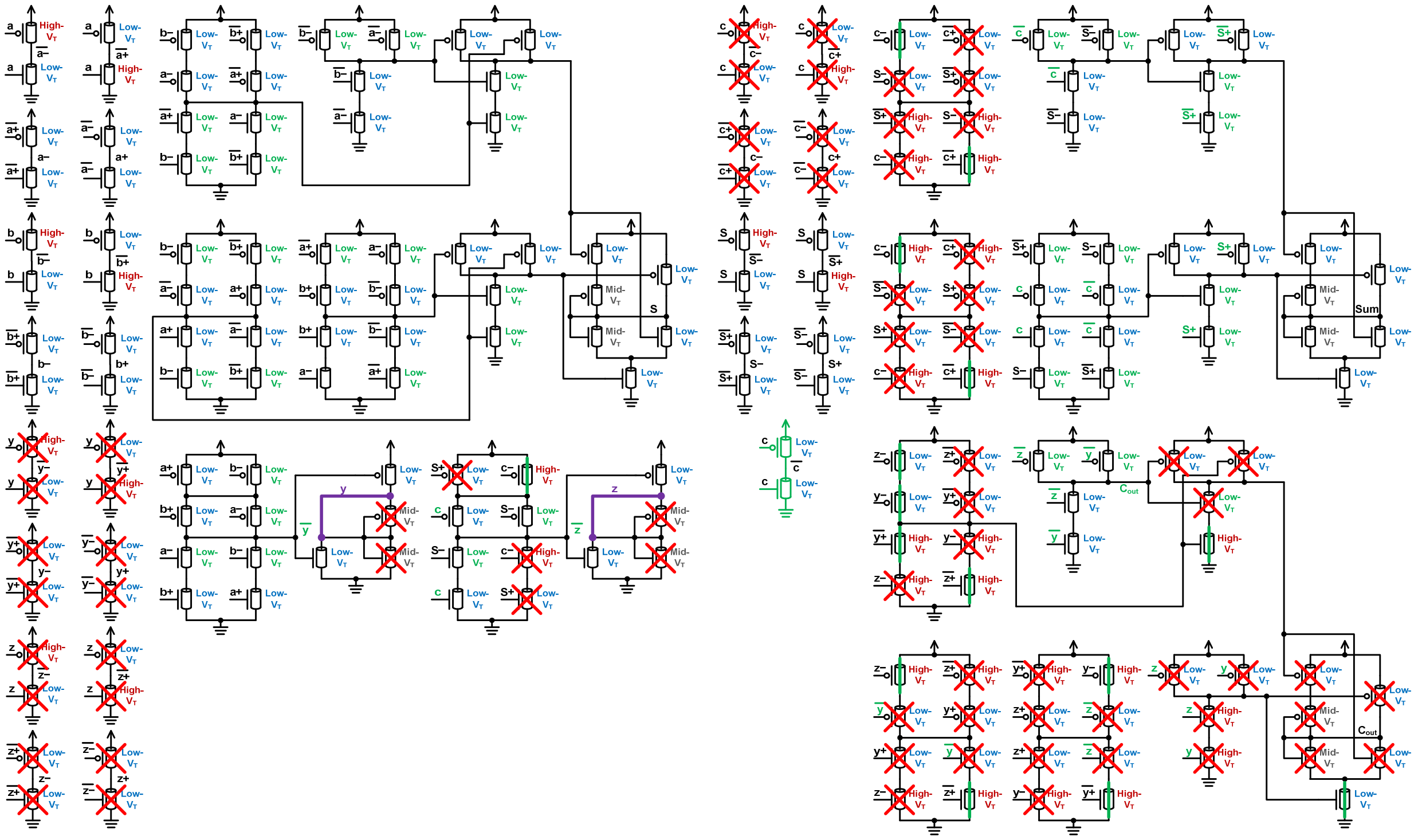

Fig. 18. The complete TFA in [17] is simplified in this paper to reach a partial TFA, where V(Carry=‘1’) = $V_{DD}$.

*J. Design in [37]*

The partial TFA in [37], whose simulation results are given in Table 15, is in its simplest form (Fig. 19), and we have not made any specific simplification to this design. However, in spite of being in the simplest form, the output *Sum* is non-full-swing in some input patterns because two n-type (p-type) transistors are placed in the PUN (PDN).

TABLE XV: Simulation Results for the TFA in [37]

| TFA | Delay (ps) | Power (μW) | PDP (fJ) | #Transistor |
|---|---|---|---|---|
| Partial [37] | 165.74 | 4.1627 | 0.6899 | 54 |

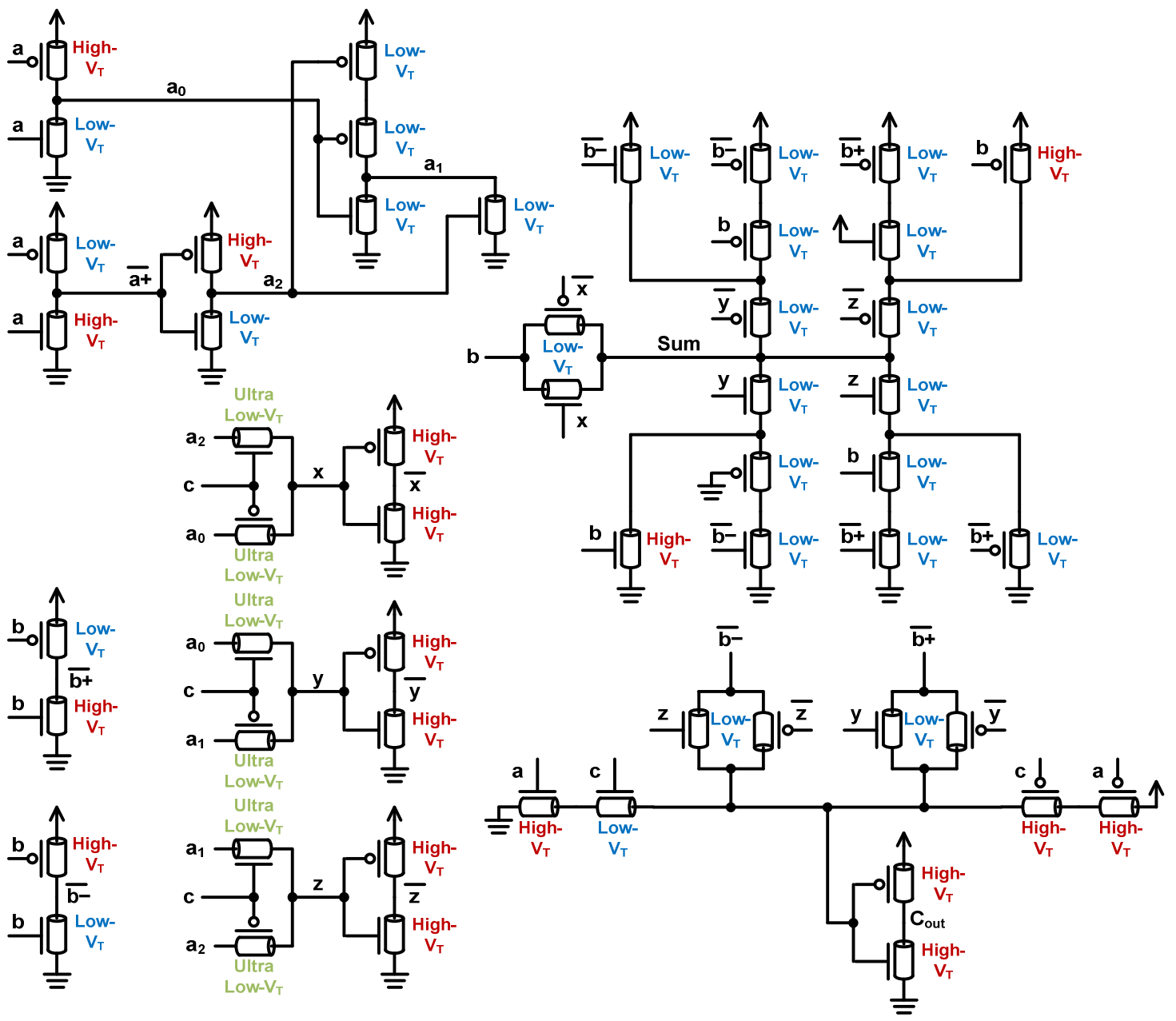

Fig. 19. The partial TFA in [37].

## IV. Simulation Results and Comparison

Synopsys HSPICE simulator is the CAD tool utilized in all of the previous papers for transistor-level circuit simulations. The same simulator and 32nm MOSFET-like CNFET library, presented at Stanford University [141], is used in this paper. This standard model includes non-idealities and has been designed for CNFETs with one or more CNT(s) as the channel material. All of the circuits are simulated in a 0.9V power supply at room temperature. Input buffers and ternary fan-out of 4 (ternary FO4) output loads also provide a standard test-bed (Fig. 20(a)), where ideal inputs pass through ternary buffers and non-ideal input signals are connected to the partial TFA under test. None of the previous papers has considered input buffers in their simulation scenarios. Additionally, the complete input pattern, including all transitions (306 ones) with the operating frequency of 1GHz is fed to the circuits (Fig. 20(b)). The maximum delay among the transitions is considered to be the maximum cell delay. Also, the average power consumption during all the transitions is the average power consumption of the cell ($P_{Average}$). The simulation results shown in the previous Section comply with the above conditions.

The chirality vectors for *High-$V_T$*, *Mid-$V_T$*, *Low-$V_T$*, and *Ultra Low-$V_T$* CNFETs are (10, 0), (14, 0), (19, 0), and (25, 0), respectively. Moreover, all CNFETs are assumed to have three CNTs under their gate terminal. This assumption is in accord with most of the previous papers. In addition, this way, the number of transistors is a close criterion for area assessment. With the aim of vivid and compact illustration, the simulation results for the newly simplified partial TFAs are shown in Table 16 once again. The simplified version of the partial TFA in [103] with the structure of DCVSL excels at delay and PDP. However, the partial TFA in [37], which has the fewest transistors, consumes the least power.

Power dissipation is a major concern in ternary designs, especially single-$V_{DD}$ ones where voltage division for the generation of ½$V_{DD}$ occurs. As indicated in (2), the total power consumed by a digital circuit ($P_{Total}$ or $P_{Average}$) is the sum of dynamic and static power ($P_{Dynamic}$ and $P_{Static}$) [142]. In (2), $a$ is the switching activity factor, $C$ is the capacitor load, and $f$ is the operating frequency. To measure static power, the amount of power dissipated while a TFA is in a stable condition (without the occurrence of any transition in the internal and output nodes) is measured by HSPICE. This measurement is repeated for all of the 18 input possibilities of the truth table of the partial TFA (Table 3), and the average amount is considered $P_{Static}$. For a deeper analysis of power consumption, the elements of power versus different operating frequencies and capacitive loads are shown in Figs. 21 and 22, respectively. The difference between total and static power implies $P_{Dynamic}$. The following are the observations from this trial:

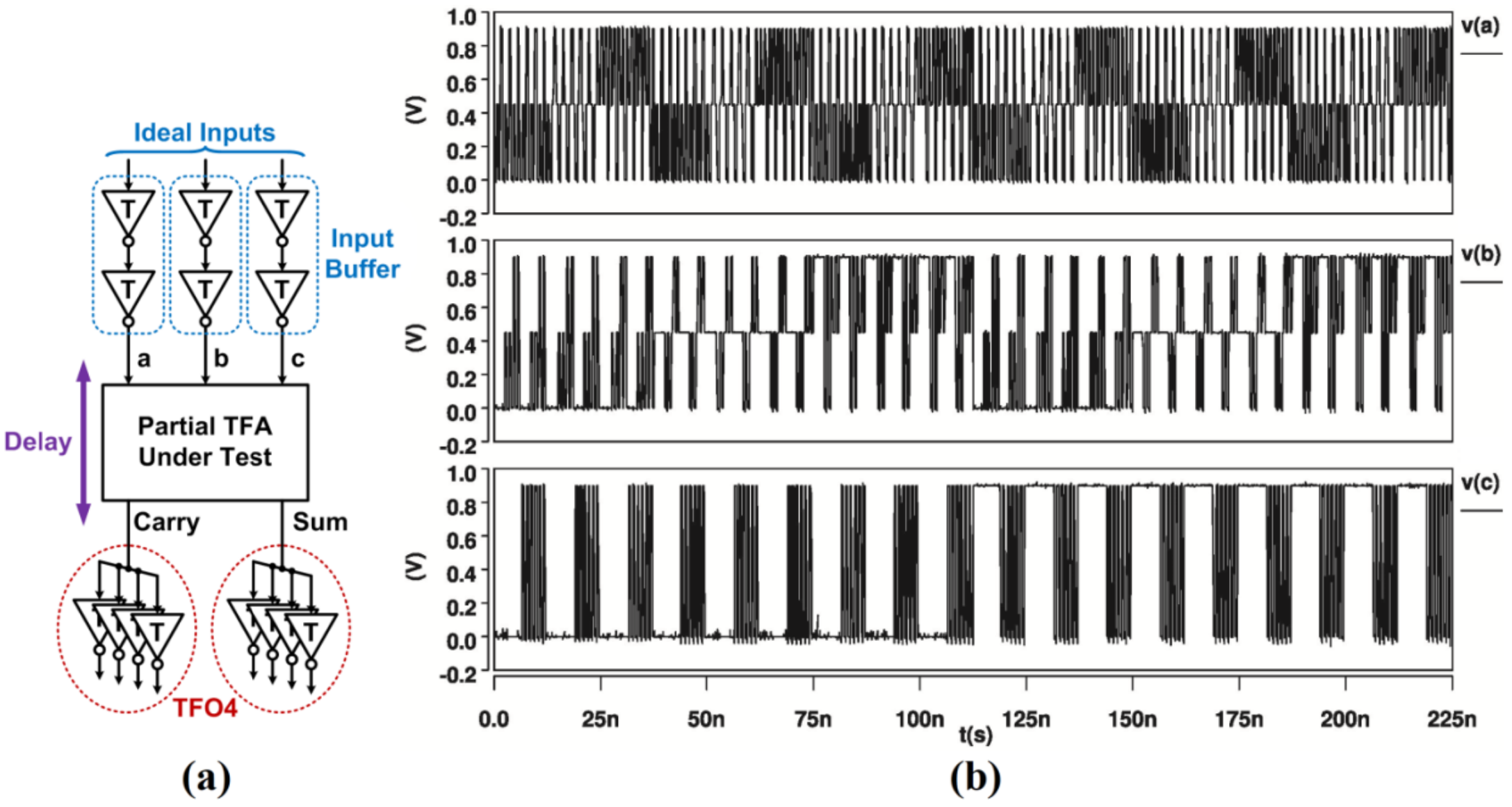

Fig. 20. Simulation setup, (a) Simulation test-bed, (b) Complete input pattern for the simulation of simplified partial TFAs.

TABLE XVI: Simulation Results of the Simplified Single-Digit Partial TFAs

| Simplified Partial TFAs of | Delay (ps) | Average Power (μW) | PDP (fJ) | #Transistor |
|---|---|---|---|---|
| [101] (Fig. 8) | 67.210 | 6.2825 | 0.4223 | 74 |
| [35] (Fig. 10) | 111.94 | 4.5080 | 0.5046 | 76 |
| [102] (Fig. 12) | 58.191 | 5.3980 | 0.3141 | 102 |
| [103] (Fig. 13) | 56.936 | 5.3516 | 0.3047 | 100 |
| [58] (Fig. 14) | 90.231 | 6.1885 | 0.5584 | 66 |
| [21] (Fig. 15) | 154.63 | 4.4792 | 0.6926 | 78 |
| [60] (Fig. 16) | 318.49 | 7.8805 | 2.5098 | 266 |
| [87] (Fig. 17) | 66.665 | 5.4544 | 0.3636 | 74 |
| [17] (Fig. 18) | 68.622 | 7.1485 | 0.4905 | 114 |
| [37]* (Fig. 19) | 165.74 | 4.1627 | 0.6899 | 54 |

* We have not made any simplification to this design.

$$P_{Total} = P_{Dynamic} + P_{Static} = a.C.f.V^2 + I_{Static}.V \quad (2)$$

- The amount of static power remains almost constant as the operating frequency and capacitive load increase. The result adheres to (2), which shows that $P_{Static}$ is independent of these factors.
- Another noticeable point is that, in most cases, static power is a significant portion of the total power consumption of TFAs.
- $P_{Dynamic}$ (and $P_{Total}$ as a result) for some of the partial TFAs increases with higher intensification than others. This phenomenon is quantified in Figs. 21 and 22 by the slopes of the curves.

Eventually, another simulation scenario for the simulation of the newly simplified partial TFAs is considered. In the second scenario, a 4-digit ripple carry adder (RCA), Fig. 23, is constructed by each of the simplified partial TFAs. Several input patterns that trigger the entire propagation path from the least significant block to the most significant block are fed to the RCA for the assessment of delay. This experiment shows that deploying {0V, $V_{DD}$} for the output carry does not adversely affect the functionality and operation of RCAs. Simulation results for this scenario can be found in Table 17. The RCA with the simplified version of the TFA in [17] excels at delay and PDP. However, the RCA with the partial TFA in [37] consumes the least power.

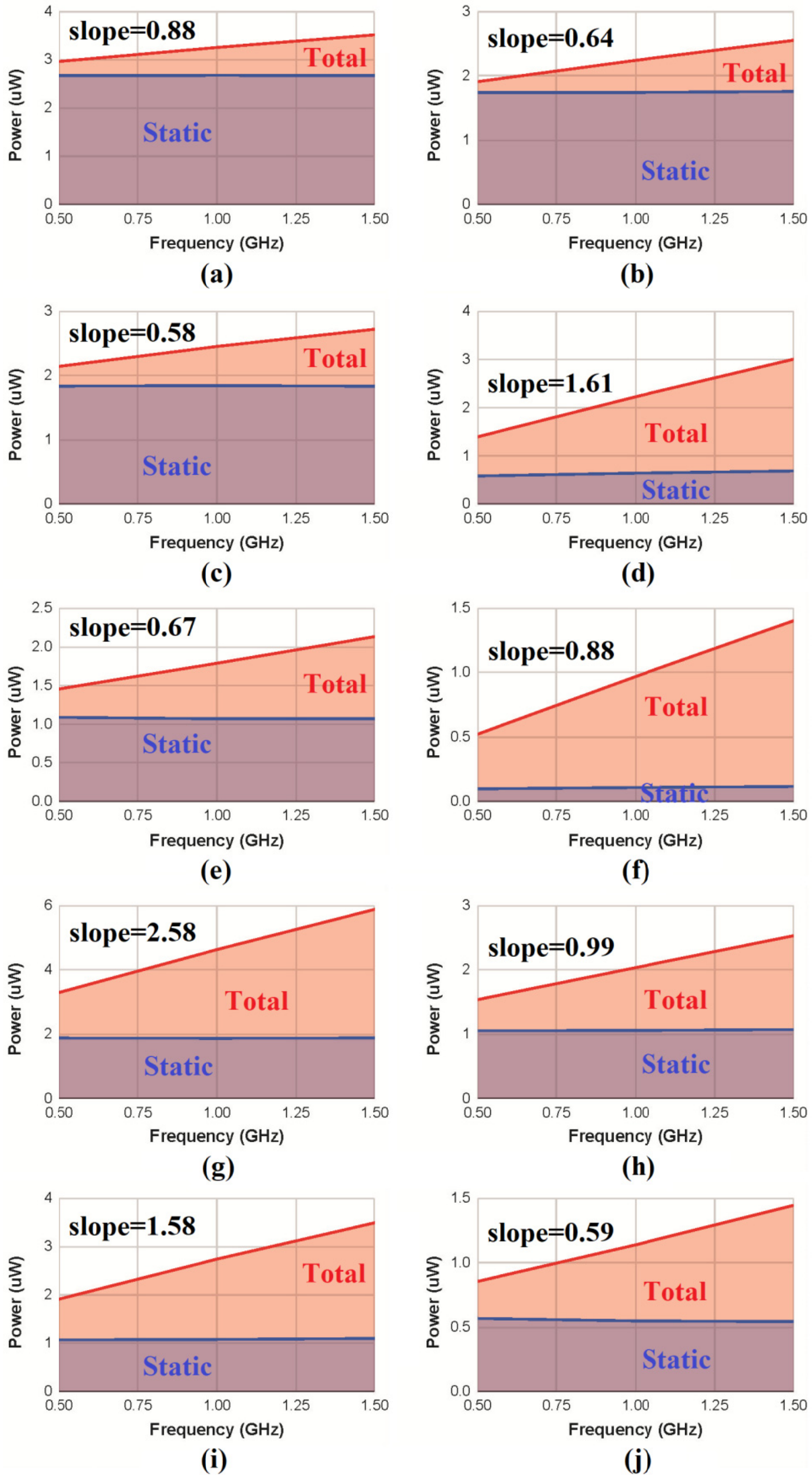

Fig. 21. Power consumption versus operating frequency, (a) [101] - Fig. 8, (b) [35] - Fig. 10, (c) [102] - Fig. 12, (d) [103] - Fig. 13, (e) [58] - Fig. 14, (f) [21] - Fig. 15, (g) [60] - Fig. 16, (h) [87] - Fig. 17, (i) [17] - Fig. 18, (j) [37] - Fig. 19.

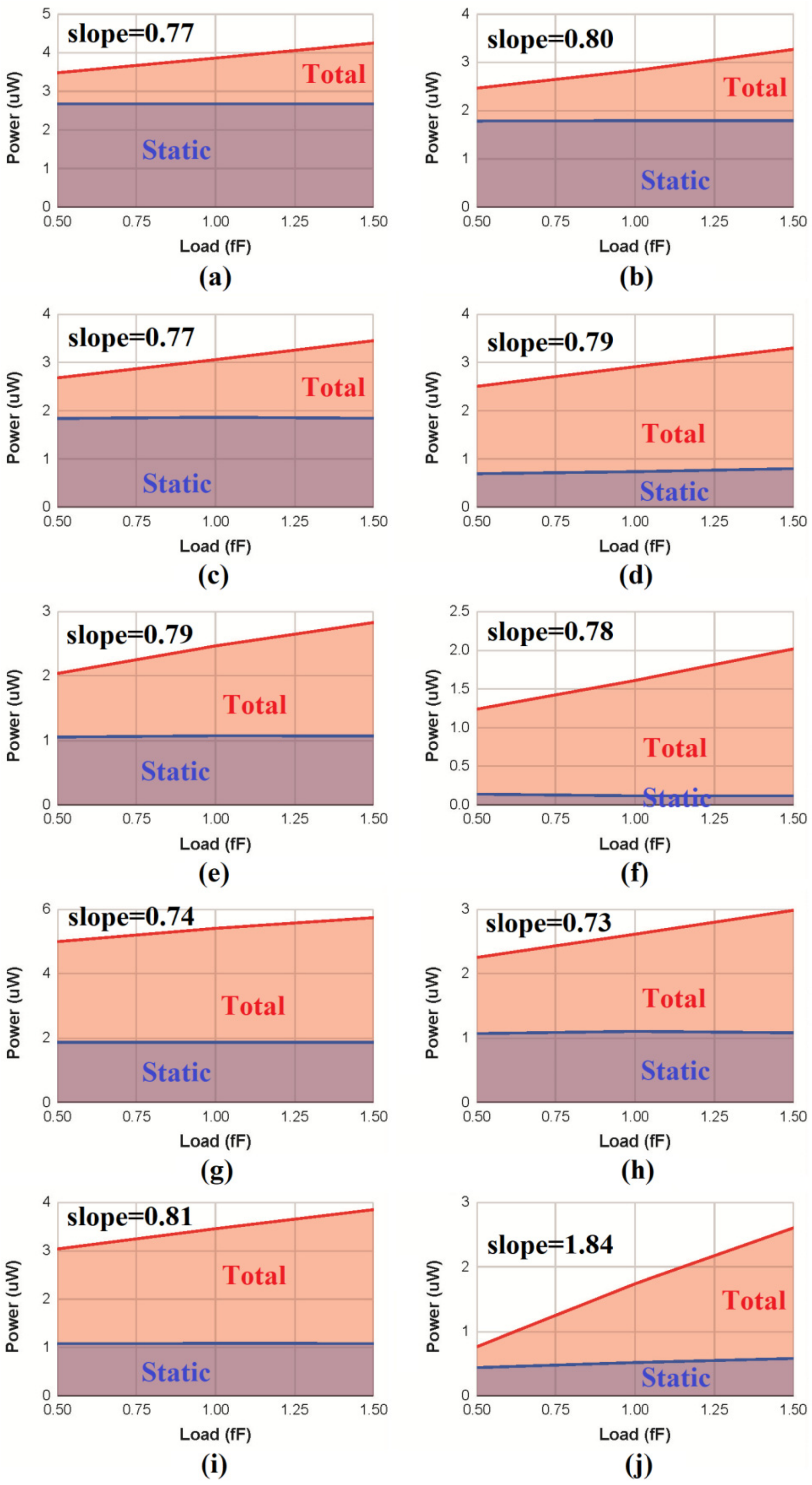

Fig. 22. Power consumption versus capacitive load, (a) [101] - Fig. 8, (b) [35] - Fig. 10, (c) [102] - Fig. 12, (d) [103] - Fig. 13, (e) [58] - Fig. 14, (f) [21] - Fig. 15, (g) [60] - Fig. 16, (h) [87] - Fig. 17, (i) [17] - Fig. 18, (j) [37] - Fig. 19.

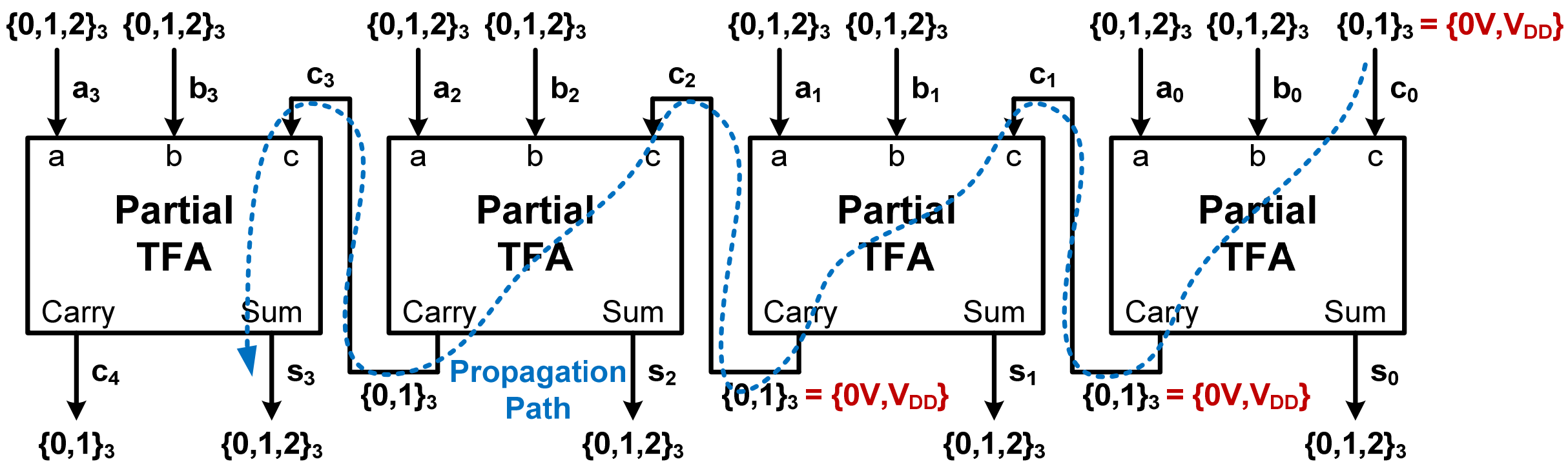

Fig. 23. A 4-digit RCA constructed by partial TFAs where V(Carry=‘1’) = $V_{DD}$.

TABLE XVII: Simulation Results of 4-Digit RCA by the Simplified Partial TFAs

| Simplified Partial TFAs of | Delay (ps) | Average Power (µW) | PDP (fJ) |
|---|---|---|---|
| [101] (Fig. 8) | 74.609 | 6.1581 | 0.4594 |
| [35] (Fig. 10) | 86.646 | 2.4320 | 0.2107 |
| [102] (Fig. 12) | 90.390 | 4.7382 | 0.4283 |
| [103] (Fig. 13) | 59.485 | 2.9451 | 0.1752 |
| [58] (Fig. 14) | 73.097 | 6.1694 | 0.4509 |
| [21] (Fig. 15) | 79.400 | 2.0073 | 0.1594 |
| [60] (Fig. 16) | 263.74 | 6.9501 | 1.8330 |
| [87] (Fig. 17) | 42.893 | 5.8932 | 0.2527 |
| [17] (Fig. 18) | 36.623 | 2.1830 | 0.0799 |
| [37]* (Fig. 19) | 143.93 | 1.8263 | 0.2629 |

* We have not made any simplification to this design.

Here are some other noticeable observations when considering all of the experiments performed in this paper:

- The designs in [21] and [102] have the same logic style unless the latter employs two binary inverters in the middle. On the one hand, they provide high driving capability and boost performance when the adder cell drives a large output load (e.g. TFO4). On the other hand, they lengthen the critical path of the cell. This is the reason why the performance of the RCA constructed by [102] is not as satisfactory/fast as when the single-digit TFA is put under test in rigorous conditions. The structure of RCA does not burden a high load on a single TFA cell as TFO4 does.
- It might seem that TFAs made up of cascaded THAs are slower than direct designs. However, the designs in [17, 101, 102] are fairly fast. Moreover, they provide a kind of parallelism in the RCA scenario when the first THAs in each bit positions can perform simultaneous computations. This is one of the reasons why the RCA constructed by [17] is very fast.
- Once the ternary inputs are decoded, the employment of ternary gates is no longer needed, and binary gates can be replaced to improve performance. The TFA in [60] and its simplified version are good examples in this regard.
- The TFA in [60], which is composed of many binary components, consumes a lot of dynamic power. In addition, with an increase in the operating frequency, $P_{Dynamic}$ increases sharply compared to other designs (Fig. 21). The main reason is that, unlike ternary logic, dynamic power is the dominant factor for power consumption in binary logic. Note that the slope in Fig. 22 is not as sharp as in Fig. 21 for this adder cell. The reason is that the capacitive loads are only added to the output nodes, *Sum* and *Carry*, whereas an increase in the operating frequency affects all of the internal nodes.

- The utilization of PTs is a helpful strategy to reduce the number of transistors, as it has led to fewer transistors in [37] and [58].
- The unbroken generation of ½$V_{DD}$ by a constant voltage division is one the reasons why the TFA in [60] has the most power dissipation. In an alternative solution, voltage division can be limited to only when the output *Sum* is supposed to be '1'. Simulation results in Figs. 21 and 22 also confirm high static power dissipation for this design.

## V. Conclusions and Future Works

A review of ternary adder cells, with a focus on TFA, has been conducted in this paper. Although the results were tabulated next to each other and one can compare the selected TFAs in this paper, we are not going to choose a specific logic family as the best option because each logic style has its advantages, disadvantages, and application. Moreover, it is neither fair nor precise to conclude from the simulation of a limited number of circuits that a logic family is superior to others. Such a firm conclusion needs far more simulations with more in-depth analyses, which can be done in future studies.

Another target of this paper was to reveal the shortcomings of the previous papers. The deficiencies exist not only in the environment where the previous TFAs have been simulated but also in how they have been designed. It has been shown in this paper that many transistors could have been eliminated from the previous TFAs by designing a partial TFA instead of a complete one. Furthermore, no voltage division is required in the *Carry* generator part. Such simplifications have been carried out in this paper, and simulation results show how efficient the simplified partial TFAs compared to their complete versions are.

Considerable efforts have been made on this topic; however, further attempts are still required to develop higher-performance ternary adders. A TFA with the faster operation, lower power consumption, and fewer transistors is needed to be considered a potential rival for the binary counterparts. It is a challenge that has not been resolved yet. Eventually, we must mention that multi-$V_{DD}$ ternary circuits, which have been excluded in this review paper, might be a possible solution. Therefore, a comprehensive survey of multi-$V_{DD}$ TFAs is worthy of conduct in the future.

## References

[1] V. Novak, "Fuzzy logic in natural language processing," IEEE Int. Conf. Fuzzy Systems, Naples, Italy, pp. 1-6, July 2017.

[2] S. Ahn, S.V. Couture, A. Cuzzocrea, K. Dam, G.M. Grasso, C.K. Leung, K.L. McCormick, and B.H. Wodi, "A fuzzy logic based machine learning tool for supporting big data business analytics in complex artificial intelligence environments," IEEE Int. Conf. Fuzzy Systems, New Orleans, USA, pp. 1-6, June 2019.

[3] E. Ozer, R. Sendag, and D. Gregg, "Multiple-valued logic buses for reducing bus energy in low-power systems," IEE Proc. Computers and Digital Techniques, vol. 153, no. 4, pp. 270-282, July 2006.

[4] F.Z. Rokhani, and G.E. Sobelman, "Multi-level signaling for energy-efficient on-chip interconnects," 7th Int. Conf. ASIC, Guilin, China, pp. 82-85, October 2007.

[5] S. Bennett, and J. Sullivan, "NAND flash memory and its place in IoT," 32nd Irish Signals and Systems Conference, Athlone, Ireland, pp. 1-6, June 2021.

[6] Z. Zilic, and K. Radecka, "Scaling and better approximating quantum Fourier transform by higher radices," IEEE Trans. Computers, vol. 56, no. 2, pp. 202-207, January 2007.

[7] V.G. Oklobdzija, Digital Design and Fabrication. Boca Raton, FL, USA: CRC Press, 2017.

[8] R. Martel, T. Schmidt, H.R. Shea, T. Hertel, and P. Avouris, "Single-and multi-wall carbon nanotube field-effect transistors," Applied Physics Letters, vol. 73, no. 17, pp. 2447-2449, October 1998.

[9] X.Y. Wang, C.T. Dong, Z.R. Wu, and Z.Q. Cheng, "A review on the design of ternary logic circuits," Chinese Physics B, vol. 30, no. 12, p. 128402, December 2021.

[10] S. Jamuna, “A brief review on multiple-valued logic-based digital circuits,” ICT with Intelligent Applications, vol. 248, pp. 329-337, 2022.
[11] A. Raychowdhury, and K. Roy, “Carbon-nanotube-based voltage-mode multiple-valued logic design,” IEEE Trans. Nanotechnology, vol. 4, no. 2, pp. 168-179, March 2005.
[12] S. Lin, Y.B. Kim, and F. Lombardi, “CNTFET-based design of ternary logic gates and arithmetic circuits,” IEEE Trans. Nanotechnology, vol. 10, no. 2, pp. 217-225, November 2009.
[13] K. Sridharan, S. Gurindagunta, and V. Pudi, “Efficient multiternary digit adder design in CNTFET technology,” IEEE Trans. Nanotechnology, vol. 12, no. 3, pp. 283-287, March 2013.
[14] S.K. Sahoo, G. Akhilesh, R. Sahoo, and M. Muglikar, “High-performance ternary adder using CNTFET,” IEEE Trans. Nanotechnology, vol. 16, no. 3, pp. 368-374, January 2017.
[15] C. Vudadha, and M.B. Srinivas, “Design of high-speed and power-efficient ternary prefix adders using CNFETs,” IEEE Trans. Nanotechnology, vol. 17, no. 4, pp. 772-782, May 2018.
[16] S. Gadgil, and C. Vudadha, “Design of CNTFET-based ternary ALU using 2:1 multiplexer based approach,” IEEE Trans. Nanotechnology, vol. 19, pp. 661-671 August 2020.
[17] S. Vidhyadharan, and S.S. Dan, “An efficient ultra-low-power and superior performance design of ternary half adder using CNFET and gate-overlap TFET devices,” IEEE Trans. Nanotechnology, vol. 20, pp. 365-376, January 2021.
[18] B. Srinivasu, and K. Sridharan, “Low-complexity multiternary digit multiplier design in CNTFET technology,” IEEE Trans. Circuits and Systems II: Express Briefs, vol. 63, no. 8, pp. 753-757, February 2016.
[19] B. Srinivasu, and K. Sridharan, “A synthesis methodology for ternary logic circuits in emerging device technologies,” IEEE Trans. Circuits and Systems I: Regular Papers, vol. 64, no. 8, pp. 2146-2159, April 2017.
[20] C. Vudadha, A. Surya, S. Agrawal, and M.B. Srinivas, “Synthesis of ternary logic circuits using 2:1 multiplexers,” IEEE Trans. Circuits and Systems I: Regular Papers, vol. 65, no. 12, pp. 4313-4325, June 2018.
[21] S. Kim, S.Y. Lee, S. Park, K.R. Kim, and S. Kang, “A logic synthesis methodology for low-power ternary logic circuits,” IEEE Trans. Circuits and Systems I: Regular Papers, vol. 67, no. 9, pp. 3138-3151, May 2020.
[22] J. Yang, H. Lee, J.H. Jeong, T. Kim, S.H. Lee, and T. Song, “Circuit-level exploration of ternary logic using memristors and MOSFETs,” IEEE Trans. Circuits and Systems I: Regular Papers, vol. 69, no. 2, pp. 707-720, February 2022.
[23] X.Y. Wang, C.T. Dong, P.F. Zhou, S.K. Nandi, S.K. Nath, R.G. Elliman, H.H.C. Iu, S.M. Kang, and J.K. Eshraghian, “Low-variance memristor-based multi-level ternary combinational logic,” IEEE Trans. Circuits and Systems I: Regular Papers, vol. 69, no. 6, pp. 2423-2434, March 2022.
[24] S. Karmakar, J.A. Chandy, and F.C. Jain, “Design of ternary logic combinational circuits based on quantum dot gate FETs,” IEEE Trans. Very Large Scale Integration (VLSI) Systems, vol. 21, no. 5, pp. 793-806, June 2012.
[25] J.L. Huber, J. Chen, J.A. McCormack, C.W. Zhou, and M.A. Reed, “An RTD/transistor switching block and its possible application in binary and ternary adders,” IEEE Trans. Electron Devices, vol. 44, no. 12, pp. 2149-2153, December 1997.
[26] N. Soliman, M.E. Fouda, A.G. Alhurbi, L.A. Said, A.H. Madian, and A.G. Radwan, “Ternary functions design using memristive threshold logic,” IEEE Access, vol. 7, pp. 48371-48381, April 2019.
[27] R.A. Jaber, A. Kassem, A.M. El-Hajj, L.A. El-Nimri, and A.M. Haidar, “High-performance and energy-efficient CNFET-based designs for ternary logic circuits,” IEEE Access, vol. 7, pp. 93871-93886, July 2019.
[28] A.D. Zarandi, M.R. Reshadinezhad, and A. Rubio, “A systematic method to design efficient ternary high performance CNTFET-based logic cells,” IEEE Access, vol. 8, pp. 58585-58593, March 2020.
[29] J.M. Aljaam, R.A. Jaber, and S.A. Al-Maadeed, “Novel ternary adder and multiplier designs without using decoders or encoders,” IEEE Access, vol. 9, pp. 56726-56735, April 2021.
[30] R.A. Jaber, J.M. Aljaam, B.N. Owaydat, S.A. Al-Maadeed, A. Kassem, and A.M. Haidar, “Ultra-low energy CNFET-based ternary combinational circuits designs,” IEEE Access, vol. 9, pp. 115951-115961, August 2021.
[31] P.C. Balla, and A. Antoniou, “Low power dissipation MOS ternary logic family,” IEEE J. Solid-State Circuits, vol. 19, no. 5, pp. 739-749, October 1984.
[32] A. Heung, and H.T. Mouftah, “Depletion/enhancement CMOS for a lower power family of three-valued logic circuits,” IEEE J. Solid-State Circuits, vol. 20, no. 2, pp. 609-616, April 1985.
[33] S. Heo, S. Kim, K. Kim, H. Lee, S.Y. Kim, Y.J. Kim, S.M., Kim, H.I. Lee, S. Lee, K.R. Kim, and S. Kang, “Ternary full adder using multi-threshold voltage graphene barristers,” IEEE Electron Device Lett., vol. 39, no. 12, pp. 1948-1951, October 2018.
[34] Z.T. Sandhie, F.U. Ahmed, and M.H. Chowdhury, “Design of ternary logic and arithmetic circuits using GNRFET,” IEEE Open J. Nanotechnology, vol. 1, pp. 77-87, September 2020.
[35] P. Keshavarzian, and R. Sarikhani, “A novel CNTFET-based ternary full adder,” Circuits, Systems, and Signal Processing, vol. 33, no. 3, pp. 665-679, March 2014.

[36] T. Sharma, and L. Kumre, "CNTFET-based design of ternary arithmetic modules," Circuits, Systems, and Signal Processing, vol. 38, no. 10, pp. 4640-4666, October 2019.
[37] S.A. Hosseini, and S. Etezadi, "A novel low-complexity and energy-efficient ternary full adder in nanoelectronics," Circuits, Systems, and Signal Processing, vol. 40, no. 3, pp. 1314-1332, March 2021.
[38] S. Tabrizchi, H. Sharifi, F. Sharifi, and K. Navi, "A novel design approach for ternary compressor cells based on CNTFETs," Circuits, Systems, and Signal Processing, vol. 35, no. 9, pp. 3310-3322, September 2016.
[39] M. Shahangian, S.A. Hosseini, and S.H. Pishgar Komleh, "Design of a multi-digit binary-to-ternary converter based on CNTFETs," Circuits, Systems, and Signal Processing, vol. 38, no. 6, pp. 2544-2563, June 2019.
[40] T. Sharma, and L. Kumre, "Energy-efficient ternary arithmetic logic unit design in CNTFET technology," Circuits, Systems, and Signal Processing, vol. 39, no. 7, pp. 3265-3288, July 2020.
[41] S. Tabrizchi, F. Sharifi, and P. Dehghani, "Energy-efficient and PVT-tolerant CNFET-based ternary full adder cell," Circuits, Systems, and Signal Processing, vol. 40, no. 7, pp. 3523-3535, July 2021.
[42] A.S. Vidhyadharan, K. Bha, and S. Vidhyadharan, "CNFET-based ultra-low-power dual-VDD V DD ternary half adder," Circuits, Systems, and Signal Processing, vol. 40, no. 8, pp. 4089-4105, August 2021.
[43] H. Samadi, A. Shahhoseini, and F. Aghaei-liavali, "A new method on designing and simulating CNTFET_based ternary gates and arithmetic circuits," Microelectronics J., vol. 63, pp. 41-48, May 2017.
[44] C. Vudadha, S.P. Parlapalli, and M.B. Srinivas, "Energy efficient design of CNFET-based multi-digit ternary adders," Microelectronics J., vol. 75, pp. 75-86, May 2018.
[45] M. Toulabinejad, M. Taheri, K. Navi, and N. Bagherzadeh, "Toward efficient implementation of basic balanced ternary arithmetic operations in CNFET technology," Microelectronics J., vol. 90, pp. 267-277, August 2019.
[46] M.Z. Jahangir, and J. Mounika, "Design and simulation of an innovative CMOS ternary 3 to 1 multiplexer and the design of ternary half adder using ternary 3 to 1 multiplexer," Microelectronics J., vol. 90, pp. 82-87, August 2019.
[47] M. Nayeri, P. Keshavarzian, and M. Nayeri, "Approach for MVL design based on armchair graphene nanoribbon field effect transistor and arithmetic circuits design," Microelectronics J., vol. 92, p. 104599, October 2019.
[48] R.A. Jaber, A.M. El-Hajj, A. Kassem, L.A. Nimri, and A.M. Haidar, "CNFET-based designs of ternary half-adder using a novel "decoder-less" ternary multiplexer based on unary operators," Microelectronics J., vol. 96, p. 104698, February 2020.
[49] A.S. Vidhyadharan, and S. Vidhyadharan, "An ultra-low-power CNFET based dual VDD ternary dynamic Half Adder," Microelectronics J., vol. 107, p. 104961, January 2021.
[50] F.M. Sardroudi, M. Habibi, and M.H. Moaiyeri, "CNFET-based design of efficient ternary half adder and 1-trit multiplier circuits using dynamic logic," Microelectronics J., vol. 113, p. 105105, July 2021.
[51] S.L. Murotiya, and A. Gupta, "Design of CNTFET-based 2-bit ternary ALU for nanoelectronics," Int. J. Electronics, vol. 101, no. 9, pp. 1244-1257, September 2014.
[52] S.L. Murotiya, and A. Gupta, "Hardware-efficient low-power 2-bit ternary ALU design in CNTFET technology," Int. J. Electronics, vol. 103, no. 5, pp. 913-927, May 2016.
[53] I.M. Salehabad, K. Navi, and M. Hosseinzadeh, "Two novel inverter-based ternary full adder cells using CNFETs for energy-efficient applications," Int. J. Electronics, vol. 107, no. 1, pp. 82-98, January 2020.
[54] A. Saha, R.K. Singh, P. Gupta, and D. Pal, "DPL-based novel CMOS 1-trit ternary full-adder," Int. J. Electronics, vol. 108, no. 2, pp. 218-236, February 2021.
[55] A.S. Vidhyadharan, and S. Vidhyadharan, "Mux based ultra-low-power ternary adders and multiplier implemented with CNFET and 45 nm MOSFETs," Int. J. Electronics, vol. 109, no. 1, pp. 1-25, January 2022.
[56] M.H. Moaiyeri, A. Doostaregan, and K. Navi, "Design of energy-efficient and robust ternary circuits for nanotechnology," IET Circuits, Devices & Systems, vol. 5, no. 4, pp .285-296, July 2011.
[57] F. Jafarzadehpour, and P. Keshavarzian, "Low-power consumption ternary full adder based on CNTFET," IET Circuits, Devices & Systems, vol. 10, no. 5, pp. 365-374, September 2016.
[58] B. Srinivasu, and K. Sridharan, "Carbon nanotube FET-based low-delay and low-power multi-digit adder designs," IET Circuits, Devices & Systems, vol. 11, no. 4, pp. 352-364, September 2017.
[59] S. Tabrizchi, A. Panahi, F. Sharifi, K. Navi, and N. Bagherzadeh, "Method for designing ternary adder cells based on CNFETs," IET Circuits, Devices & Systems, vol. 11, no. 5, pp. 465-470, September 2017.
[60] B.D. Madhuri, and S. Sunithamani, "Design of ternary logic gates and circuits using GNRFETs," IET Circuits, Devices & Systems, vol. 14, no. 7, pp. 972-979, November 2020.
[61] H. Gundersen, and Y. Berg, "A novel balanced ternary adder using recharged semi-floating gate devices," IEEE 36th Int. Symp. Multiple-Valued Logic, Singapore, pp. 18-18, May 2006.

[62] H. Gundersen, and Y. Berg, "Fast addition using balanced ternary counters designed with CMOS semi-floating gate devices," IEEE 37th Int. Symp. Multiple-Valued Logic, Oslo, Norway, pp. 30-30, May 2007.
[63] J. Yang, H. Lee, J.H. Jeong, T.H. Kim, S.H. Lee, and T. Song, "A practical implementation of the ternary logic using memristors and MOSFETs," IEEE 51st Int. Symp. Multiple-Valued Logic, Nur-sultan, Kazakhstanm, pp. 183-188, May 2021.
[64] J. Ko, K. Park, S. Yong, T. Jeong, T.H. Kim, and T. Song, "An optimal design methodology of ternary logic in ISO-device ternary CMOS," IEEE 51st Int. Symp. Multiple-Valued Logic, Nur-sultan, Kazakhstanm, pp. 189-194, May 2021.
[65] R.F. Mirzaee, M.H. Moaiyeri, M. Maleknejad, K. Navi, and O. Hashemipour, "Dramatically low-transistor-count high-speed ternary adders," IEEE 43rd Int. Symp. Multiple-Valued Logic, Toyama, Japan, pp. 170-175, May 2013.
[66] M. Moradi, R.F. Mirzaee, and K. Navi, "Ternary versus binary multiplication with current-mode CNTFET-based K-valued converters," IEEE 46th Int. Symp. Multiple-Valued Logic, Sapporo, Japan, pp. 17-22, May 2016.
[67] Y. Kang, J. Kim, S. Kim, S. Shin, E.S. Jang, J.W. Jeong, K.R. Kim, and S. Kang, "A novel ternary multiplier based on ternary CMOS compact model," IEEE 47th Int. Symp. Multiple-Valued Logic, Novi Sad, Serbia, pp. 25-30, May 2017.
[68] C.K. Vudadha, and M.B. Srinivas, "Design methodologies for ternary logic circuits," IEEE 48th Int. Symp. Multiple-Valued Logic, Linz, Austria, pp. 192-197, May 2018.
[69] S. Kim, S.Y. Lee, S. Park, and S. Kang, "Design of quad-edge-triggered sequential logic circuits for ternary logic," IEEE 49th Int. Symp. Multiple-Valued Logic, Fredericton, Canada, pp. 37-42, May 2019.
[70] H. Lee, H. Jang, J. Yun, H. Jin, J. Kim, Y. Kim, and T. Song, T., "Ternary competitive to binary: A novel implementation of ternary logic using depletion-mode and conventional MOSFETs," IEEE 52nd Int. Symp. Multiple-Valued Logic, Dallas, USA, pp. 21-26, May 2022.
[71] J. Kim, Y. Kim, H. Lee, J. Yun, H. Jang, H. Jin, J. Park, B. Kim, and T. Song, "A convenient implementation of the ternary logic: Using anti-ambipolar transistors and PMOS based on printed carbon nanotubes," IEEE 52nd Int. Symp. Multiple-Valued Logic, Dallas, USA, pp. 15-20, May 2022.
[72] R.P. Hallworth, F.G. Heath, "Semiconductor circuits for ternary logic," Proc. IEE-Part C: Monographs, vol. 109, no. 15, pp. 219-225, March 1962.
[73] I. Halpern, and M. Yoeli, "Ternary arithmetic unit," Proc. IEE, vol. 115, no. 10, pp. 1385-1388, October 1968.
[74] G.V. Anjaneyulu, "A new approach to the design of ternary adder," IETE J. Research, vol. 18, no. 7, pp. 323-324, July 1972.
[75] P.G. Sankar, "A novel ternary half adder & one bit multiplier circuits based on emerging sub-32nm FET technology," Int. Conf. Intelligent Computing and Communication for Smart World, Erode, India, pp. 198-203, December 2018.
[76] M. Nayeri, P. Keshavarzian, and M. Nayeri, "High-speed ternary half adder based on GNRFET," J. Nanoanalysis, vol. 6, no. 3, pp. 193-198, September 2019.
[77] A. Mohammaden, M.E. Fouda, L.A. Said, and A.G. Radwan, "Memristor-CNTFET based ternary full adders," IEEE 63rd Int. Midwest Symp. Circuits and Systems, Springfield, USA, pp. 562-565, August 2020.
[78] F. Zahoor, F.A. Hussin, F.A. Khanday, M.R. Ahmad, I. Mohd Nawi, C.Y. Ooi, and F.Z. Rokhani, "Carbon nanotube field effect transistor (CNFET) and resistive random access memory (RRAM) based ternary combinational logic circuits," Electronics, vol. 10, no. 1, p.79, January 2021.
[79] P. Srikanth, B. Srinivasu, and N. Kaushik, "Ternary full adder in CMOS-memristor technology," IEEE 22nd Int. Conf. Nanotechnology, Palma de Mallorca, Spain, pp. 89-92, July 2022.
[80] M. Lin, Q. Han, W. Luo, X. Wang, J. Chen, and W. Lyu, "A ternary memristor full adder based on literal operation and module operation," Int. J. Circuit Theory and Applications, vol. 50, no. 8, pp. 2932-2940, August 2022.
[81] L. Li, Z. Zhang, and C. Chen, "An area-efficient ternary full adder using hybrid SET-MOS technology," IEEE 17th Int. Conf. Nanotechnology, Pittsburgh, USA, pp. 576-578, July 2017.
[82] S.M. Ghadamgahi, R. Sabbaghi-Nadooshan, and K. Navi, "Physical proof and design of ternary full adder circuit in ternary quantum-dot cellular automata technology," Int. J. Numerical Modelling: Electronic Networks, Devices and Fields, p. e2995, February 2022.
[83] Y. Ji, S. Chang, H. Wang, Q. Huang, J. He, and F. Yi, "Multi-valued logic design methodology with double negative differential resistance transistors," Micro & Nano Lett., vol. 12, no. 10, pp. 738-743, October 2017.
[84] A. Srivastava, and K. Venkatapathy, "Design and implementation of a low power ternary full adder," VLSI Design, vol. 4, no. 1, pp. 75-81, 1996.
[85] H. Gundersen, "On the potential of CMOS recharged semi-floating gate devices used in balanced ternary logic," 17th Int. Workshop Post-Binary ULSI Systems, Dallas, USA, pp. 1-8, May 2008.
[86] S. Kim, T. Lim, and S. Kang, "An optimal gate design for the synthesis of ternary logic circuits," 23rd Asia and South Pacific Design Automation Conf., Jeju, South Korea, pp. 476-481, January 2018.

[87] D. Etiemble, "Best CNTFET ternary adders?", arXiv preprint arXiv:2101.01516, January 2021.
[88] R. Mariani, R. Roncella, R. Saletti, and P. Terreni, "On the realisation of delay-insensitive asynchronous circuits with CMOS ternary logic," 3rd Int. Symp. Advanced Research in Asynchronous Circuits and Systems, Eindhoven, Netherlands, pp. 54-62, April 1997.
[89] S.L. Murotiya, A. Gupta, and S. Vasishth, "CNTFET-based design of dynamic ternary full adder cell," Annual IEEE India Conf., Pune, India, pp. 1-5, December 2014.
[90] S. Rezaie, R.F. Mirzaee, K. Navi, and O. Hashemipour, "From static ternary adders to high-performance race-free dynamic ones," The J. Engineering, vol. 2015, no. 12, pp. 371-382, December 2015.
[91] S.M. Ghadamgahi, R. Sabbaghi-Nadooshan, and K. Navi, "Novel ternary adders and subtractors in quantum cellular automata," The J. Supercomputing, vol. 78, pp. 18454–18496, June 2022.
[92] F.M. Sardroudi, M. Habibi, and M.H. Moaiyeri, "A low-power dynamic ternary full adder using carbon nanotube field-effect transistors," AEU-Int. J. Electronics and Communications, vol. 131, p. 153600, March 2021.
[93] N.I. Chernov, N.N. Prokopenko, and N.V. Butyrlagin, "Method of analog-to-digital conversion of current signals conveyed by sensor based on multi-valued adder," 2nd Int. Ural Conf. Measurements, Chelyabinsk, Russia, pp. 74-80, October 2017.
[94] N. Butyrlagin, N. Chernov, N. Prokopenko, and V. Yugai, "Design of two-valued and multivalued current digital adders based on the mathematical tool of linear algebra," IEEE East-West Design & Test Symp., Kazan, Russia, pp. 1-6, September 2018.
[95] A.G. Asibelagh, and R.F. Mirzaee, "Partial ternary full adder versus complete ternary full adder. Int. Conf. Electrical, Communication, and Computer Engineering, Istanbul, Turkey, pp. 1-6, June 2020.
[96] S. Firouzi, S. Tabrizchi, F. Sharifi, and A.H. Badawy, "High performance, variation-tolerant CNFET ternary full adder a process, voltage, and temperature variation-resilient design," Computers & Electrical Engineering, vol. 77, pp. 205-216, July 2019.
[97] A. Srivastava, and K. Venkatapathy, "Performance of CMOS ternary full adder at liquid nitrogen temperature," Cryogenics, vol. 35, no. 9, pp. 599-605, September 1995.
[98] A.P. Dhande, and V.T. Ingole, "Design and implementation of 2 bit ternary ALU slice," 3rd Int. Conf. Sciences of Electronic, Technologies of Information and Telecommunications, Tunisia, pp. 1-11, March 2005.
[99] X. Zeng, and P. Wang, "Design of low-power complementary pass-transistor and ternary adder based on multi-valued switch-signal theory," 8th Int. Conf. ASIC, Changsha, China, pp. 851-854, October 2009.
[100] M.H. Moaiyeri, R.F. Mirzaee, K. Navi, and O. Hashemipour, "Efficient CNTFET-based ternary full adder cells for nanoelectronics," Nano-Micro Lett., vol. 3, no. 1, pp. 43-50, March 2011.
[101] S.A. Ebrahimi, P. Keshavarzian, S. Sorouri, and M. Shahsavari, "Low power CNTFET-based ternary full adder cell for nanoelectronics," Int. J. Soft Computing and Engineering, vol. 2, no. 2, pp. 291-295, May 2012.
[102] R.F. Mirzaee, K. Navi, and N. Bagherzadeh, "High-efficient circuits for ternary addition," VLSI Design, vol. 2014, article ID 534587, pp. 1-15, 2014.
[103] S.L. Murotiya, and A. Gupta, "A novel design of ternary full adder using CNTFETs," Arabian J. Science and Engineering, vol. 39, no. 11, pp. 7839-7846, November 2014.
[104] S.K. Sahoo, G. Akhilesh, and R. Sahoo, "Design of a high performance carry generation circuit for ternary full adder using CNTFET," IEEE Int. Symp. Nanoelectronic and Information Systems, Bhopal, India, pp. 46-49, December 2017.
[105] S.L. Murotiya, and A. Gupta, "Design of high speed ternary full adder and three-input XOR circuits using CNTFETs," 28th Int. Conf. VLSI Design, Bangalore, India, pp. 292-297, January 2015.
[106] L.S. Phanindra, M.N. Rajath, V. Rakesh, and K.V. Patel, "A novel design and implementation of multi-valued logic arithmetic full adder circuit using CNTFET," IEEE Int. Conf. Recent Trends in Electronics, Information & Communication Technology, Bangalore, India, pp. 563-568, May 2016.
[107] V. Sridevi, and T. Jayanthy, "Minimization of CNTFET ternary combinational circuits using negation of literals technique," Arabian J. Science and Engineering, vol. 39, no. 6, pp. 4875-4890, June 2014.
[108] S. Rani, and B. Singh, "CNTFET based 4-trit hybrid ternary adder-subtractor for low power & high-speed applications," Silicon, vol. 14, no. 2, pp. 689-702, January 2022.
[109] P.V. Saidutt, V. Srinivas, P.S. Phaneendra, and N.M. Muthukrishnan, "Design of encoder for ternary logic circuits," Asia Pacific Conf. Postgraduate Research in Microelectronics and Electronics, Hyderabad, India, pp. 85-88, December 2012.
[110] M. Maleknejad, R.F. Mirzaee, K. Navi, and O. Hashemipour, "Multi-Vt ternary circuits by carbon nanotube field effect transistor technology for low-voltage and low-power applications," J. Computational and Theoretical Nanoscience, vol. 11, no. 1, pp. 110-118, January 2014.

[111] M. Maeen, and K. Navi, “Design and evaluation of an efficient carbon nano-tube field effect transistor-based ternary full adder cell for nanotechnology,” J. Computational and Theoretical Nanoscience, vol. 11, no. 9, pp. 1934-1941, September 2014.

[112] R.F. Mirzaee, and K. Navi, “Optimized adder cells for ternary ripple-carry addition,” IEICE Trans. Information and systems, vol. 97, no. 9, pp. 2312-2319, September 2014.

[113] K.M. Jameel, “Experimental Design of a Ternary Full Adder using Pseudo N-type Carbon Nano tube FETs,” Int. Research J. Engineering and Technology, vol. 2, no. 9, pp. 2395-0056, December 2015.

[114] M.H. Moaiyeri, M. Nasiri, and N. Khastoo, “An efficient ternary serial adder based on carbon nanotube FETs,” Engineering Science and Technology, an Int. J., vol. 19, no. 1, pp. 271-278, March 2016.

[115] M. Muglikar, R. Sahoo, and S.K. Sahoo, “High performance ternary adder using CNTFET,” 3rd Int. Conf. Devices, Circuits and Systems, Coimbatore, India, pp. 236-239, March 2016.

[116] M. Bastami, and R.F. Mirzaee, “Integration of CTL, PTL, and DCVSL for designing a novel fast ternary half adder,” The CSI J. Computer Science and Engineering, vol. 15, no. 1, pp. 15-21, 2017.

[117] N.H. Bastani, M.H. Moaiyeri, and K. Navi, “Carbon nanotube field effect transistor switching logic for designing efficient ternary arithmetic circuits,” J. Nanoelectronics and Optoelectronics, vol. 12, no. 2, pp. 118-129, February 2017.

[118] M.M. Ghanatghestani, B. Ghavami, and H. Pedram, “A ternary full adder cell based on carbon nanotube FET for high-speed arithmetic units,” J. Nanoelectronics and Optoelectronics, vol. 13, no. 3, pp. 368-377, March 2018.

[119] E. Shahrom, and S.A. Hosseini, “A new low power multiplexer based ternary multiplier using CNTFETs,” AEU-Int. J. Electronics and Communications, vol. 93, pp. 191-207, September 2018.

[120] F. Sharifi, A. Panahi, M.H. Moaiyeri, H. Sharifi, and K. Navi, “High performance CNFET-based ternary full adders,” IETE J. Research, vol. 64, no. 1, pp. 108-115, January 2018.

[121] M. Huang, X. Wang, G. Zhao, P. Coquet, and B. Tay, “Design and implementation of ternary logic integrated circuits by using novel two-dimensional materials,” Applied Sciences, vol. 9, no. 20, p. 4212, 2019.

[122] J.L. Merlin, T.A. Khan, and T.S. Hameed, “Design of a low power three bit ternary prefix adder using CNTFET technology. AIP Conf. Proceedings, vol. 2222, no. 1, p. 020005, April 2020.

[123] M.S. Ghoneim, A. Mohammaden, R. Hesham, and A.H. Madian, “Low power scalable ternary hybrid full adder realization,” 32nd Int. Conf. Microelectronics, Aqaba, Jordan, pp. 1-4, December 2020.

[124] R.A. Jaber, B. Owaidat, A. Kassem, and A.M. Haidar, “A novel low-energy CNTFET-based ternary half-adder design using unary operators,” Int. Conf. Innovation and Intelligence for Informatics, Computing and Technologies, Sakheer, Bahrain, pp. 1-6, December 2020.

[125] E. Nikbakht, D. Dideban, and N. Moezi, “A half adder design based on ternary multiplexers in carbon nano-tube field effect transistor (CNFET) technology,” ECS J. Solid State Science and Technology, vol. 9, no. 8, p. 081001, September 2020.

[126] N. Dehabadi, and R.F. Mirzaee, “Ternary DCVS half adder with built-in boosters,” J. Intelligent Procedures in Electrical Technology, vol. 11, no. 42, pp. 41-56, 2020.

[127] N. Vejendla, P. Jamanchipalli, S. Bontha, J. Dendeti, B. Bolloju, and K.K. Kuppili, “Design of low PDP ternary circuits utilizing carbon nanotube field-effect transistors,” Intelligent Computing in Control and Communication, vol. 702, pp. 247-265, January 2021.

[128] Y. Pendashteh, and S.A. Hosseini, “Novel low-complexity and energy-efficient fuzzy min and max circuits in nanoelectronics,” AEU-Int. J. Electronics and Communications, vol. 138, p. 153858, August 2021.

[129] R.A. Jaber, H. Bazzi, A. Haidar, B. Owaidat, and A. Kassem, “1-trit Ternary Multiplier and Adder Designs Using Ternary Multiplexers and Unary Operators,” Int. Conf. Innovation and Intelligence for Informatics, Computing, and Technologies, Zallaq, Bahrain, pp. 292-297, September 2021.

[130] T. Sharma, and L. Kumre, “Efficient ternary compressor design using capacitive threshold logic in CNTFET technology,” IETE J. Research, vol. 14, pp.1-11, January 2021.

[131] U. Panwar, and P. Sharma, “Design and implementation of ternary adder for High-Performance arithmetic applications by using CNTFET material,” Materials Today: Proceedings, vol. 63, pp. 773-777, June 2022.

[132] M. Yousefi, K. Monfaredi, and Z. Moradi, “Design and simulation of pseudo ternary adder based on CNTFET,” AUT J. Electrical Engineering, vol. 54, no. 2, pp. 361-376, December 2022.

[133] A. Saha, R. Pal, T. Kumari, R.K. Singh, S. Chakraborty, and J. Ghosh, “Fast complete ternary addition with novel 3:1 T-multiplexer,” Micro and Nanosystems, vol. 14, no. 4, pp. 304-313, December 2022.

[134] A. Mohammaden, M.E. Fouda, I. Alouani, L.A. Said, and A.G. Radwan, “CNTFET-based ternary multiply-and-accumulate unit,” Electronics, vol. 11, no. 9, p. 1455, January 2022.

[135] S.V. RatanKumar, L.K. Rao, and M.K. Kumar, "Design of ternary logic circuits using pseudo n-type CNTFETs," ECS J. Solid State Science and Technology, vol. 11, no. 11, p. 111003, November 2022.
[136] J. Liang, L. Chen, J. Han, and F. Lombardi, "Design and evaluation of multiple valued logic gates using pseudo N-type carbon nanotube FETs," IEEE Trans. Nanotechnology, vol. 13, no. 4, pp. 695-708, April 2014.
[137] S. Rezaie, R.F. Mirzaee, K. Navi, and O. Hashemipour, "New dynamic ternary minimum and maximum circuits with reduced switching activity and without any additional voltage sources," Int. J. High Performance Systems Architecture, vol. 5, no. 3, pp. 153-165, 2015.
[138] R.F. Mirzaee, T. Nikoubin, K. Navi, and O. Hashemipour, "Differential cascode voltage switch (DCVS) strategies by CNTFET technology for standard ternary logic," Microelectronics J., vol. 44, no. 12, pp. 1238-1250, December 2013.
[139] N. Azimi, R.F. Mirzaee, K. Navi, and A.M. Rahmani, "Ternary DDCVSL: a combined dynamic logic style for standard ternary logic with single power source," IET Computers & Digital Techniques, vol. 14, no. 4, pp. 166-175, June 2020.
[140] S. Waser, and M. J. Flynn, Introduction to Arithmetic for Digital Systems Designers, Oxford University Press: London, 1995.
[141] Stanford University CNFET Model website, 2008, Available at: https://nano.stanford.edu/model.php
[142] N. Weste, and D. Harris, "CMOS VLSI design: A circuit and systems perspective," 4th Edition, Addison-Wesley, Boston, 2011.